# Propositional computability logic II

Giorgi Japaridze[*]

Department of Computing Sciences, Villanova University, 800 Lancaster Avenue, Villanova, PA 19085, USA.
Email: giorgi.japaridze@villanova.edu URL: http://www.csc.villanova.edu/~japaridz/

**Abstract**

Computability logic is a formal theory of computational tasks and resources. Its formulas represent interactive computational problems, logical operators stand for operations on computational problems, and validity of a formula is understood as being a scheme of problems that always have algorithmic solutions. The earlier article "Propositional computability logic I" proved soundness and completeness for the (in a sense) minimal nontrivial fragment **CL1** of computability logic. The present paper extends that result to the significantly more expressive propositional system **CL2**. What makes **CL2** more expressive than **CL1** is the presence of two sorts of atoms in its language: *elementary atoms*, representing elementary computational problems (i.e. predicates), and *general atoms*, representing arbitrary computational problems. **CL2** conservatively extends **CL1**, with the latter being nothing but the general-atom-free fragment of the former.

*MSC*: primary: 03B47; secondary: 03F50; 03B70; 68Q10; 68T27; 68T30; 91A05

*Keywords*: Computability logic; Interactive computation; Game semantics; Linear logic; Constructive logics

## 1 Introduction

Being a continuation of [3], this article fully relies on the terminology, notation, conventions and technical results of its predecessor, with which the reader is assumed to be familiar.

The atoms of our old friend **CL1** represent predicates rather than computational problems in general. Hence, **CL1** only describes valid computability principles for elementary problems. This is a rather severe limitation of expressive power. For example, from the provability of $p \vee \neg p$, the soundness of **CL1** only allows us to conclude that, for every predicate $p$, the elementary problem $p \vee \neg p$ is computable, i.e. true — the fact which is already known from classical logic. Well, of course, in the same way we can also discover a series of less known facts, such as, say, that for any predicates $p$ and $q$, the problems $(p \sqcap q) \vee \neg(p \sqcap q)$, $(p \sqcup (q \wedge p)) \vee \neg(p \sqcup (q \wedge p))$, etc. are computable (and furthermore, due to the constructive character of the soundness theorem, we would also know how, exactly, to compute such problems). Yet, this would not be sufficient for us to find that every problem of the form $A \vee \neg A$ is computable. Such a conclusion could not be automatically made even if we managed to show — reasoning in the metatheory of **CL1** rather than **CL1** itself, of course — that every **CL1**-formula of the form $F \vee \neg F$ is provable. Instead, we would only know that $A \vee \neg A$ is computable as long as $A$ is a $(\sqcap, \sqcup, \neg, \wedge, \vee, \rightarrow)$-combination of elementary problems. While the class of problems of this sort is certainly interesting and nontrivial, it is only a modest fraction of the collection of all entities that we call interactive computational problems.

By simply redefining the semantics of the language of **CL1** and no longer requiring that its atoms be interpreted as elementary problems, we would certainly gain a lot. But perhaps just as much would be lost: the class of valid formulas would shrink, victimizing many innocent principles such as, say, $p \rightarrow p \wedge p$ or $p \vee p \rightarrow p \sqcup p$. The point is that elementary problems are meaningful and interesting in their own right, and losing the ability to differentiate them from problems in general would be too much of a sacrifice. Needless to mention, we would have to say "goodbye" to the nice fact that classical logic is a fragment of the new logic.

---

[*]This material is based upon work supported by the National Science Foundation under Grant No. 0208816

Computability logic has a better solution. It simply allows two sorts of atoms in its language, one for elementary problems and the other for all problems. This way, not only do we have the ability to characterize valid principles for problems of either sort within the same formal system, but we can as well capture principles that intermix elementary problems with not-necessarily-elementary ones. Such an approach also has technical advantages. As we are going to see, logic **CL2**, whose language extends that of **CL1** by adding to it the second sort of atoms, has a rather simple (yet unusual) axiomatization, while it remains unclear and perhaps questionable whether there is a reasonable axiomatization for the fragment of computability logic whose language only has the second sort of atoms.

This article is devoted to a soundness and completeness proof for the above-mentioned system **CL2**. Its main result, with a forward reference to the present paper (and without any proof), has been announced in [4].

## 2  Logic CL2

The language of **CL2** is the same as that of **CL1**, with the only "little" difference that, along with the old atoms of **CL1** which we now call **elementary**, it has an additional sort of nonlogical atoms called **general**. We continue using the lowercase letters $p, q, r, s$ as metavariables for elementary atoms, and will be using the uppercase $P, Q, R, S$ as metavariables for general atoms. We refer to the well-formed expressions of this language as **CL2-formulas**, or simply **formulas**. They are built from atoms in the standard way using the connectives $\neg, \wedge, \vee, \rightarrow, \sqcap, \sqcup$. The formulas that do not contain elementary nonlogical atoms we call **general-base**, and the formulas that do not contain general atoms (i.e. simply **CL1**-formulas) we call **elementary-base**. This terminology also extends to the corresponding two fragments of **CL2**; in particular, the *general-base fragment* of **CL2** is the set of all general-base theorems of **CL2**, and the *elementary-base fragment* of **CL2** is the set of all elementary-base theorems of **CL2**. In view of the promised soundness and completeness of **CL2**, we may guess that the elementary-base fragment of **CL2** is nothing but (the set of theorems of) **CL1**.

An **interpretation** for the language of **CL2** a function that sends each nonlogical elementary atom (as before) to an elementary game, and sends each general atom to any, not-necessarily-elementary, static game. This mapping extends to all formulas in the same way as in the case of **CL1**. Many of our old concepts such as validity, uniform validity, surface occurrence and positive/negative occurrence, straightforwardly extend to this new language and there is no need to redefine them. An **elementary formula** also means the same as before: this is a formula of classical propositional logic, i.e. a formula that does not contain general atoms, $\sqcap$ and $\sqcup$. The **elementarization** of a **CL2**-formula $F$ means the result of replacing in $F$ every surface occurrence of the form $G_1 \sqcap \ldots \sqcap G_n$ by $\top$, every surface occurrence of the form $G_1 \sqcup \ldots \sqcup G_n$ by $\bot$ *and*, in addition, replacing every positive surface occurrence of each general atom by $\bot$ and every negative surface occurrence of each general atom by $\top$. Finally, as before, a formula is said to be **stable** iff its elementarization is a classical tautology; otherwise it is **instable**.

The rules of inference of **CL2** are the two rules **(a)** and **(b)** of **CL1** — which are now applied to any **CL2**-formulas rather than just elementary-base formulas — plus the following additional rule:

**Rule (c):** $F' \vdash F$, where $F'$ is the result of replacing in $F$ two — one positive and one negative — surface occurrences of some general atom by a nonlogical elementary atom that does not occur in $F$.

**Example 2.1** The following is a **CL2**-proof of $P \wedge P \rightarrow P$:
1. $p \wedge P \rightarrow p$   (from $\emptyset$ by Rule **(a)**);
2. $P \wedge P \rightarrow P$   (from 1 by Rule **(c)**).

On the other hand, while **CL2** certainly proves $p \rightarrow p \wedge p$, it does not prove $P \rightarrow P \wedge P$. Indeed, this formula is instable and does not contain $\sqcap$ or $\sqcup$, so it cannot be derived by Rules **(a)** or **(b)**. If it is derived by Rule **(c)**, the premise should be $p \rightarrow P \wedge p$ or $p \rightarrow p \wedge P$ for some nonlogical elementary atom $p$. In either case we deal with an instable formula that contains no choice operators and only has one occurrence of a general atom, so that it cannot be derived by any of the three rules of **CL2**.

**Exercise 2.2** Verify that:
1. $\mathbf{CL2} \vdash P \vee \neg P$
2. $\mathbf{CL2} \not\vdash P \sqcup \neg P$
3. $\mathbf{CL2} \vdash P \rightarrow P \sqcap P$
4. $\mathbf{CL2} \vdash (P \wedge Q) \vee (R \wedge S) \rightarrow (P \vee R) \wedge (Q \vee S)$ (Blass's principle)[1]
5. $\mathbf{CL2} \vdash p \wedge (p \rightarrow Q) \wedge (p \rightarrow R) \rightarrow Q \wedge R$
6. $\mathbf{CL2} \not\vdash P \wedge (P \rightarrow Q) \wedge (P \rightarrow R) \rightarrow Q \wedge R$
7. $\mathbf{CL2} \vdash P \sqcap (Q \vee R) \rightarrow (P \sqcap Q) \vee (P \sqcap R)$
8. $\mathbf{CL2} \not\vdash (P \sqcap Q) \vee (P \sqcap R) \rightarrow P \sqcap (Q \vee R)$
9. $\mathbf{CL2} \vdash (p \sqcap Q) \vee (p \sqcap R) \rightarrow p \sqcap (Q \vee R)$

As we remember, **CL1** is a conservative extension of classical logic. **CL2**, in turn, is a conservative extension of **CL1**. This fact, of course, is implied by our soundness/completeness theorems. But it can as well be seen directly through a simple syntactic analysis, taking into account that Rule **(c)** introduces a general atom that never disappears in later formulas of a **CL2**-proof.

Below comes our main theorem. It is simply the later-proven Lemmas 4.1 and 5.1 put together:

**Theorem 2.3** $\mathbf{CL2} \vdash F$ *iff $F$ is valid (any* **CL2***-formula $F$). Furthermore:*

*a) There is an effective procedure that takes a* **CL2***-proof of an arbitrary formula $F$ and constructs an HPM $\mathcal{H}$ such that, for every interpretation $*$, $\mathcal{H}$ computes $F^*$.*

*b) If $\mathbf{CL2} \not\vdash F$, then $F^*$ is not computable for some interpretation $*$ that interprets all elementary atoms of $F$ as finitary predicates of arithmetical complexity $\Delta_2$, and interprets all general atoms of $F$ as problems of the form $(A_1^1 \sqcup \ldots \sqcup A_m^1) \sqcap \ldots \sqcap (A_1^m \sqcup \ldots \sqcup A_m^m)$, where each $A_i^j$ is a finitary predicate of arithmetical complexity $\Delta_2$.*

**CL2** is almost the full propositional fragment of the first-order logic **FD** introduced in [2], where the latter was conjectured (Conjecture 25.4) to be sound and complete with respect to computability-logic semantics. What is missing in **CL2** is the logical general atom $, interpreted as a 'computational problem of universal strength'. In order to keep our proofs shorter, we have not included $ in the language. However, extending our soundness/completeness proof in a way that accommodates $ does not present a serious challenge. Anyway, **FD** can be easily seen to be a conservative extension of **CL2**, which means that our Theorem 2.3 yields a positive verification of Conjecture 25.4 of [2] restricted to the language of **CL2**.

Next, reasoning as in [3] (Section 5, Theorem 5.10), Theorem 2.3 allows us to find that Conjecture 26.2 of [2], restricted to the language of **CL2**, is also correct. That is, we have:

**Theorem 2.4** *A* **CL2***-formula is valid if and only if it is uniformly valid.*

Likewise, our Theorem 2.3 implies a positive verification of the correspondingly restricted Conjecture 24.4 of [2] as well. The latter sounds as follows:

> *If a formula $F$ is not valid, then $F^*$ is not computable for some interpretation $*$ that interprets all atoms as finitary, determined, strict, unistructural problems.*

---

[1]This formula is a *binary tautology*, meaning a tautology of classical propositional logic where each nonlogical atom occurs at most twice. Blass [1] showed that the set of binary tautologies and their substitutional instances was precisely the multiplicative fragment of the logic induced by his game semantics for linear logic. It is not hard to see that the $\sqcap, \sqcup$-free subfragment of the general-base fragment of **CL2** yields exactly the same class of formulas. The fact that the two, technically rather different semantics (see Section 27 of [2] for a discussion of the differences), introduced with different motivations, validate same formulas is certainly a positive sign, signifying that both of the semantics are natural. Whether such equivalence extends to the full general-base fragment of **CL2** is unknown as no axiomatic/syntactic characterization has been found so far for the full additive-multiplicative propositional fragment of the logic induced by Blass's semantics.

Two of the above game properties have not been (properly) defined in [3], so they need to be explained here. A *strict* game is a game where, in every legal position, at most one of the players has legal moves. And a *determined* game is a game where, on any fixed input, one of the players has a winning strategy, even though not necessarily an algorithmic one. In precise terms, non-algorithmic strategies can be understood as HPMs with oracles; such machines, discussed in Section 18 of [2], generalize ordinary HPMs in the same way as oracle Turing machines generalize ordinary Turing machines. The above-quoted statement of Conjecture 24.4 of [2] is an immediate consequence of clause (b) of our Theorem 2.3 and the fact — known from [2] — that $\sqcap, \sqcup$-combinations of finitary predicates are finitary, determined, strict and unistructural.

It is also worth noting that **CL2**, just like **CL1**, is decidable, with a brute force decision algorithm obviously running in at most polynomial space. Whether there are more efficient algorithms is unknown.

## 3  Technical preliminaries

The rest of this paper is devoted to a proof of Theorem 2.3. This section contains some necessary preliminaries.

### 3.1  Hyperformulas

In the bottom-up (from conclusion to premises) view, Rule **(c)** introduces two occurrences of some new nonlogical elementary atom. For technical convenience, we want to differentiate elementary atoms introduced this way from all other elementary atoms, and also to somehow keep track of the exact origin of each such elementary atom $q$ — that is, remember what general atom $P$ was replaced by $q$ when Rule **(c)** was applied. For this purpose, we extend the language of **CL2** by adding to it a new sort of non-logical atoms, called **hybrid**. In particular, each hybrid atom is a pair consisting of a general atom $P$, called its **general component**, and a nonlogical elementary atom $q$, called its **elementary component**. We denote such a pair by $P_q$. It is assumed that, for every nonlogical elementary atom $q$ and every general atom $P$, the language has the (unique) hybrid atom $P_q$. As we are going to see later, the presence of $P_q$ in a (modified **CL2**-) proof will be an indication of the fact that, in the bottom-up view of proofs, $q$ has been introduced by Rule **(c)** and that when this happened, the general atom that $q$ replaced was $P$.

What we call **hyperformulas** are defined in the same way as **CL2**-formulas, with the only difference that now atomic expressions can be of any of the three (elementary, general or hybrid) sorts. "*Subhyperformula*" in the context of hyperformulas means the same as "subformula" in the context of formulas.

As in the case of formulas, by a **surface occurrence** of a subexpression in a given hyperformula $F$ we mean an occurrence that is not in the scope of $\sqcap$ and/or $\sqcup$. Understanding $G \to H$ as an abbreviation for $\neg G \vee H$, an occurrence of a subexpression in a hyperformula is **positive** (resp. **negative**) iff it is in the scope of an even (resp. odd) number of occurrences of $\neg$. An **elementary hyperformula** is one not containing $\sqcap$ and $\sqcup$, as well as general and hybrid atoms. Thus, 'elementary hyperformula' and 'elementary **CL2**-formula' (as well as 'elementary **CL1**-formula', as well as 'formula of classical propositional logic') mean the same. The **elementarization**
$$\|F\|$$
of a hyperformula $F$ is the result of replacing, in $F$, every surface occurrence of the form $G_1 \sqcap \ldots \sqcap G_n$ by $\top$, every surface occurrence of the form $G_1 \sqcup \ldots \sqcup G_n$ by $\bot$, every positive (resp. negative) surface occurrence of each general atom by $\bot$ (resp. $\top$), and every surface occurrence of each hybrid atom by the elementary component of that atom. As in the case of formulas, we say that a hyperformula $F$ is **stable** iff $\|F\|$ is valid in the classical sense; otherwise it is **instable**.

A hyperformula $F$ is said to be **balanced** iff, for every hybrid atom $P_q$ occurring in $F$, the following two conditions are satisfied:

1. $F$ has exactly two occurrences of $P_q$, where one occurrence is positive and the other occurrence is negative, and both occurrences are surface occurrences;

2. the elementary atom $q$ does not occur in $F$, nor is it the elementary component of any hybrid atom occurring in $F$ other than $P_q$.

In our soundness proof for **CL2** we will employ a "version" of **CL2** called **CL2°**. Unlike **CL2** whose language consists only of formulas, the language of **CL2°** allows any balanced hyperformulas. The rules of **CL2°** are Rules **(a)** and **(b)** of **CL2** (only now applied to any balanced hyperformulas rather than just **CL2**-formulas) plus the following Rule **(c°)** instead of the old Rule **(c)**:

**Rule (c°):** $F' \vdash F$, where $F$ is the result of replacing in $F'$ both occurrences of some hybrid atom $P_q$ by its general component $P$.

**Lemma 3.1** *For any **CL2**-formula $G$, if $\mathbf{CL2} \vdash G$, then $\mathbf{CL2°} \vdash G$.*

*Furthermore, there is an effective procedure that converts any **CL2**-proof of any formula $G$ into a **CL2°**-proof of $G$.*

**Proof.** Consider any **CL2**-proof tree for $G$, i.e. a tree every node of which is labeled with a **CL2**-formula that follows by one of the rules of **CL2** from the set of (the labels of) its children, with $G$ being the label of the root. By abuse of terminology, here we identify the nodes of this tree with their labels, even though, of course, it may be the case that different nodes have same labels. For each node $F$ of the tree that is derived from its child $F'$ by Rule **(c)** — in particular, where $F'$ is the result of replacing in $F$ a positive and a negative surface occurrences of a general atom $P$ by a nonlogical elementary atom $q$ — do the following: replace all (both) occurrences of $q$ by the hybrid atom $P_q$ in $F'$ as well as in all of its descendants in the tree. It is not hard to see that this way we will get a **CL2°**-proof of $G$. That the resulting tree is indeed a **CL2°**-proof formally can be verified by induction on the hight of the **CL2**-proof tree. □

By the *general dehibridization* of a hyperformula $F$ we mean the **CL2**-formula that results from $F$ by replacing in the latter every hybrid atom by its general component. Where $*$ is an interpretation and $F$ is a hyperformula, we define the game

$$F^*$$

as $G^*$, where $G$ is the general dehibridization of $F$. Extending the earlier-established lingo to hyperformulas, for a hyperformula $F$ and an interpretation $*$, whenever $F^* = A$, we say that $*$ **interprets** $F$ as $A$.

## 3.2 Perfect interpretations

A game $A$ is said to be **constant** iff it does not depend on input, i.e., for every two inputs $e_1$ and $e_2$, we have $\mathbf{Wn}^A_{e_1} = \mathbf{Wn}^A_{e_2}$ and $\mathbf{Lr}^A_{e_1} = \mathbf{Lr}^A_{e_2}$. Of course, for such a game $A$, there is no difference between "legal" and "unilegal", and we can always write $\mathbf{LR}^A$ instead of $\mathbf{Lr}^A_e$; likewise, since $e$ is irrelevant, we can safely omit this parameter in $\mathbf{Wn}^A_e$ and simply write $\mathbf{Wn}^A$ instead. For a constant game $A$, we will say "$\Gamma$ is a legal run (position) of $A$" to mean that $\Gamma \in \mathbf{LR}^A$, and say "$\Gamma$ is a $\wp$-won run of $A$" to mean that $\mathbf{Wn}^A\langle\Gamma\rangle = \wp$; similarly, we will just say "$\wp$-illegal..." to mean "$\wp$-illegal ... on some (= all) $e$".

For a game $A$ and input $e$, the *e-instantiation* of $A$, denoted

$$e[A],$$

is defined by stipulating that, for every input $f$, $\mathbf{Wn}^{e[A]}_f = \mathbf{Wn}^A_e$ and $\mathbf{Lr}^{e[A]}_f = \mathbf{Lr}^A_e$. Thus, the game $e[A]$ does not depend on input $f$. Intuitively, $e[A]$ is the constant game obtained from $A$ by fixing the input $e$ for it once and forever. Note that when $A$ is a constant game, for any input $e$, we have $e[A] = A$. Based on the definitions of our game operations, it is also easy to see that we always have $e[\neg A] = \neg e[A]$, $e[A \to B] = e[A] \to e[B]$, $e[A_1 \wedge \ldots \wedge A_n] = e[A_1] \wedge \ldots \wedge e[A_n]$, and similarly for $\vee, \sqcap, \sqcup$.

An interpretation $*$ is said to be **perfect** iff it interprets every atom as a constant game. All of our game operations preserve the constant property of games ([2], Theorem 14.1), which means that perfect interpretations interpret all (hyper)formulas as constant games. This fact may be worth marking as we will often implicitly rely on it. For an interpretation $*$ and input $e$, the **perfect interpretation induced by** $(*, e)$ is the interpretation $\dagger$ that interprets each (elementary or general) atom $L$ as the constant game $e[L^*]$.

**Lemma 3.2** *Assume $F$ is any hyperformula, $e$ any input, $*$ any interpretation and $\dagger$ the perfect interpretation induced by $(*, e)$. Then $e[F^*] = F^\dagger$.*

**Proof.** Induction on the complexity of $F$. For an atomic $F$, $e[F^*] = F^\dagger$ is immediate. And the inductive step is also straightforward, taking into account that the operations $e[\ldots], *, \dagger$ commute with $\neg, \wedge, \vee, \rightarrow, \sqcap, \sqcup$. □

## 3.3 Prefixation lemmas

Throughout this paper we follow the notational convention established in [2, 3, 4], according to which letter $\wp$ exclusively ranges over players, lowercase Greek letters range over moves, and uppercase Greek letters range over runs, with $\Phi, \Psi, \Theta, \Omega$ typically used for finite runs (positions), and $\Gamma, \Delta$ for any runs.

Let us also take a note of one technicality. Even though the **Lr** subclauses of the official definition of $\neg, \wedge, \vee$ given in [3] (Definition 3.2) involve a nonempty position $\Phi$, by condition (a) of Definition 3.1 of [3] it is clear that those subclauses automatically extend to any run $\Gamma$. We will often implicitly rely on this fact.

Remember the operation of $\Phi$-prefixation ([3], Definition 3.6), the result of applying which to a game $A$ is denoted $\langle\Phi\rangle A$. According to definition, $\langle\Phi\rangle A$ is defined iff $\Phi$ is a unilegal position of $A$. For readability and compactness of formulations, let us agree on the following:

**Convention 3.3** Every time we make a statement that involves an expression "$\langle\Phi\rangle A$", unless otherwise specified, we imply that $\Phi$ is a unilegal position of game $A$ and hence $\langle\Phi\rangle A$ is defined.

The expression $\langle\Psi\rangle\langle\Phi\rangle A$ below and elsewhere should be read as $\langle\Psi\rangle(\langle\Phi\rangle A)$.

**Lemma 3.4** *For any game $A$ and positions $\Phi, \Psi$, we have:*
  *1. If $\langle\Phi,\Psi\rangle A$ is defined, then so is $\langle\Psi\rangle\langle\Phi\rangle A$, and vice versa.*
  *2. When $\langle\Phi,\Psi\rangle A$ — or, equivalently, $\langle\Psi\rangle\langle\Phi\rangle A$ — is defined, we have $\langle\Phi,\Psi\rangle A = \langle\Psi\rangle\langle\Phi\rangle A$.*

**Proof.** *Clause 1.* Suppose $\langle\Phi,\Psi\rangle A$ is defined, i.e. $\langle\Phi,\Psi\rangle \in \mathbf{LR}^A$. Then, by condition (a) of Definition 3.1 of [3], $\Phi \in \mathbf{LR}^A$. This means that $\langle\Phi\rangle A$ is defined; next, by the definition of prefixation, $\Psi \in \mathbf{LR}^{\langle\Phi\rangle A}$ iff $\langle\Phi,\Psi\rangle \in \mathbf{LR}^A$, so that we have $\Psi \in \mathbf{LR}^{\langle\Phi\rangle A}$, i.e. $\langle\Psi\rangle\langle\Phi\rangle A$ is (also) defined. The 'vice versa' part can be handled in a similar/symmetric way.

*Clause 2.* Consider any input $e$ and any run $\Gamma$. We want to show that $\Gamma \in \mathbf{Lr}_e^{\langle\Psi\rangle\langle\Phi\rangle A}$ iff $\Gamma \in \mathbf{Lr}_e^{\langle\Phi,\Psi\rangle A}$, and $\mathbf{Wn}_e^{\langle\Psi\rangle\langle\Phi\rangle A}\langle\Gamma\rangle = \mathbf{Wn}_e^{\langle\Phi,\Psi\rangle A}\langle\Gamma\rangle$.

By the definition of prefixation, $\Gamma \in \mathbf{Lr}_e^{\langle\Psi\rangle\langle\Phi\rangle A}$ iff $\langle\Psi,\Gamma\rangle \in \mathbf{Lr}_e^{\langle\Phi\rangle A}$. In turn, $\langle\Psi,\Gamma\rangle \in \mathbf{Lr}_e^{\langle\Phi\rangle A}$ iff $\langle\Phi,\Psi,\Gamma\rangle \in \mathbf{Lr}_e^A$. Finally, $\langle\Phi,\Psi,\Gamma\rangle \in \mathbf{Lr}_e^A$ iff $\Gamma \in \mathbf{Lr}_e^{\langle\Phi,\Psi\rangle A}$.

Next, again immediately from the definition of prefixation, we have: $\mathbf{Wn}_e^{\langle\Psi\rangle\langle\Phi\rangle A}\langle\Gamma\rangle = \mathbf{Wn}_e^{\langle\Phi\rangle A}\langle\Psi,\Gamma\rangle = \mathbf{Wn}_e^A\langle\Phi,\Psi,\Gamma\rangle = \mathbf{Wn}_e^{\langle\Phi,\Psi\rangle A}\langle\Gamma\rangle$. □

In [3] we used the notation $\bar{\wp}$ for "$\wp$'s adversary", and the notation $\bar{\Gamma}$ for the result of changing every label $\wp$ to $\bar{\wp}$ (and vice versa) in run $\Gamma$. In this paper we go back to the notation established in [2], and write

$$\neg\wp \text{ and } \neg\Gamma$$

instead of $\bar{\wp}$ and $\bar{\Gamma}$, respectively.

**Lemma 3.5** *For any game $A$ and position $\Phi$ with $\Phi \in \mathbf{LR}^{\neg A}$ — or, equivalently, $\neg\Phi \in \mathbf{LR}^A$ — we have $\langle\Phi\rangle\neg A = \neg(\langle\neg\Phi\rangle A)$.*

**Proof.** First of all, note that, by the definition of (the game operation) $\neg$, the conditions $\Phi \in \mathbf{LR}^{\neg A}$ and $\neg\Phi \in \mathbf{LR}^A$ are indeed equivalent. Assume $\Phi \in \mathbf{LR}^{\neg A}$. Consider any input $e$ and any run $\Gamma$. We want to show that $\Gamma \in \mathbf{Lr}_e^{\langle\Phi\rangle\neg A}$ iff $\Gamma \in \mathbf{Lr}_e^{\neg(\langle\neg\Phi\rangle A)}$, and $\mathbf{Wn}_e^{\langle\Phi\rangle\neg A}\langle\Gamma\rangle = \mathbf{Wn}_e^{\neg(\langle\neg\Phi\rangle A)}\langle\Gamma\rangle$.

By the definition of prefixation, $\Gamma \in \mathbf{Lr}_e^{\langle\Phi\rangle\neg A}$ iff $\langle\Phi,\Gamma\rangle \in \mathbf{Lr}_e^{\neg A}$. In turn, by the definition of $\neg$, $\langle\Phi,\Gamma\rangle \in \mathbf{Lr}_e^{\neg A}$ iff $\neg\langle\Phi,\Gamma\rangle \in \mathbf{Lr}_e^A$. Of course, $\neg\langle\Phi,\Gamma\rangle = \langle\neg\Phi, \neg\Gamma\rangle$. Thus we get: $\Gamma \in \mathbf{Lr}_e^{\langle\Phi\rangle\neg A}$ iff $\langle\neg\Phi,\neg\Gamma\rangle \in \mathbf{Lr}_e^A$. Again by the definition of prefixation, $\langle\neg\Phi,\neg\Gamma\rangle \in \mathbf{Lr}_e^A$ iff $\neg\Gamma \in \mathbf{Lr}_e^{\langle\neg\Phi\rangle A}$. And, again by the definition of $\neg$, $\neg\Gamma \in \mathbf{Lr}_e^{\langle\neg\Phi\rangle A}$ iff $\Gamma \in \mathbf{Lr}_e^{\neg(\langle\neg\Phi\rangle A)}$. Thus, $\Gamma \in \mathbf{Lr}_e^{\langle\Phi\rangle\neg A}$ iff $\Gamma \in \mathbf{Lr}_e^{\neg(\langle\neg\Phi\rangle A)}$.

Next, excluding for safety the trivial case $\Gamma \notin \mathbf{LR}^{\langle\Phi\rangle \neg A}$ which — in view of condition (c) of Definition 3.1 of [3] — is taken care of by the previous paragraph, we have:

$\mathbf{Wn}_e^{\langle\Phi\rangle \neg A}\langle\Gamma\rangle = \mathbf{Wn}_e^{\neg A}\langle\Phi, \Gamma\rangle$ (by the definition of prefixation);
$\mathbf{Wn}_e^{\neg A}\langle\Phi, \Gamma\rangle = \neg\mathbf{Wn}_e^{A}\langle\neg\Phi, \neg\Gamma\rangle$ (by the definition of $\neg$);
$\neg\mathbf{Wn}_e^{A}\langle\neg\Phi, \neg\Gamma\rangle = \neg\mathbf{Wn}_e^{\langle\neg\Phi\rangle A}\langle\neg\Gamma\rangle$ (by the definition of prefixation);
$\neg\mathbf{Wn}_e^{\langle\neg\Phi\rangle A}\langle\neg\Gamma\rangle = \mathbf{Wn}_e^{\neg(\langle\neg\Phi\rangle A)}\langle\Gamma\rangle$ (by the definition of $\neg$).

This chain of equations yields the desired $\mathbf{Wn}_e^{\langle\Phi\rangle \neg A}\langle\Gamma\rangle = \mathbf{Wn}_e^{\neg(\langle\neg\Phi\rangle A)}\langle\Gamma\rangle$. $\square$

Remember the notation

$$\Gamma^\gamma$$

that we started using in [3], meaning the result of removing from $\Gamma$ all labeled moves except those of the form $\wp\gamma\beta$, and then deleting the prefix $\gamma$ in the remaining moves, i.e. replacing each such $\wp\gamma\beta$ by $\wp\beta$.

**Lemma 3.6** *For any games $A_1, \ldots, A_n$ ($n \geq 2$) and unilegal position $\Phi$ of $A_1 \vee \ldots \vee A_n$, we have*

$$\langle\Phi\rangle(A_1 \vee \ldots \vee A_n) = \langle\Phi^{1.}\rangle A_1 \vee \ldots \vee \langle\Phi^{n.}\rangle A_n.$$

**Proof.** Assume $\Phi \in \mathbf{LR}^{A_1 \vee \ldots \vee A_n}$. Note that then, by the definition of $\vee$, each $\Phi^{i.}$ is in $\mathbf{LR}^{A_i}$ and thus the $\langle\Phi^{i.}\rangle A_i$ are defined.

We prove the above equality by induction on the length of $\Phi$. The basis case with $\Phi = \langle\rangle$ is straightforward. Now suppose $\Phi = \langle\Psi, \wp\gamma\rangle$, so that

$$\langle\Phi\rangle(A_1 \vee \ldots \vee A_n) = \langle\Psi, \wp\gamma\rangle(A_1 \vee \ldots \vee A_n). \tag{1}$$

By Lemma 3.4,

$$\langle\Psi, \wp\gamma\rangle(A_1 \vee \ldots \vee A_n) = \langle\wp\gamma\rangle\langle\Psi\rangle(A_1 \vee \ldots \vee A_n). \tag{2}$$

Since $\Phi$ is a unilegal position of $A_1 \vee \ldots \vee A_n$, so is $\Psi$, and, by the induction hypothesis, $\langle\Psi\rangle(A_1 \vee \ldots \vee A_n) = \langle\Psi^{1.}\rangle A_1 \vee \ldots \vee \langle\Psi^{n.}\rangle A_n$. Hence

$$\langle\wp\gamma\rangle\langle\Psi\rangle(A_1 \vee \ldots \vee A_n) = \langle\wp\gamma\rangle(\langle\Psi^{1.}\rangle A_1 \vee \ldots \vee \langle\Psi^{n.}\rangle A_n). \tag{3}$$

By clause 3(a) of Lemma 3.7 of [3], $\gamma = i.\beta$, where $i \in \{1, \ldots, n\}$ and $\langle\wp\beta\rangle \in \mathbf{LR}^{\langle\Psi^{i.}\rangle A_i}$. Without loss of generality, let us assume here that $i = 1$. Then, by clause 3(b) of the same lemma,

$$\langle\wp\gamma\rangle(\langle\Psi^{1.}\rangle A_1 \vee \ldots \vee \langle\Psi^{n.}\rangle A_n) = (\langle\wp\beta\rangle\langle\Psi^{1.}\rangle A_1) \vee \langle\Psi^{2.}\rangle A_2 \vee \ldots \vee \langle\Psi^{n.}\rangle A_n. \tag{4}$$

By Lemma 3.4, $\langle\wp\beta\rangle\langle\Psi^{1.}\rangle A_1 = \langle\Psi^{1.}, \wp\beta\rangle A_1$. In turn, we clearly have $\langle\Psi^{1.}, \wp\beta\rangle = \langle\Psi, \wp 1.\beta\rangle^{1.}$, i.e. $\langle\Psi^{1.}, \wp\beta\rangle = \langle\Psi, \wp\gamma\rangle^{1.}$, i.e. $\langle\Psi^{1.}, \wp\beta\rangle = \langle\Phi\rangle^{1.}$. It is also obvious that, for every $j \neq 1$, we have $\Psi^{j.} = \langle\Psi, \wp 1.\beta\rangle^{j.}$, i.e. $\Psi^{j.} = \Phi^{j.}$. Therefore,

$$(\langle\wp\beta\rangle\langle\Psi^{1.}\rangle A_1) \vee \langle\Psi^{2.}\rangle A_2 \vee \ldots \vee \langle\Psi^{n.}\rangle A_n = \langle\Phi^{1.}\rangle A_1 \vee \langle\Phi^{2.}\rangle A_2 \vee \ldots \vee \langle\Phi^{n.}\rangle A_n. \tag{5}$$

Now, the chain (1)-(5) of equations yields the desired $\langle\Phi\rangle(A_1 \vee \ldots \vee A_n) = \langle\Phi^{1.}\rangle A_1 \vee \ldots \vee \langle\Phi^{n.}\rangle A_n$. $\square$

By a *choice hyperformula* we mean a non-atomic hyperformula whose main operator is $\sqcap$ or $\sqcup$. A **quasiatom** of a hyperformula $E$ is a surface occurrence of a subhyperformula in $E$ that is either an atom (of any of the three sorts) or a choice hyperformula. Note that a quasiatom in not just a subhyperformula but rather a subhyperformula together with a particular occurrence. E.g., in $P \wedge \neg P$, the two different occurrences of $P$ present two different quasiatoms. However, for readability (and by abuse of terminology), we will usually identify a quasiatom with the corresponding hyperformula $G$, and simply say "the quasiatom $G$" once it is clear from the context which of the possibly many occurrences of $G$ we mean.

A quasiatom $G$ of a hyperformula $F$ is said to be **positive** (resp. **negative**) (in $F$) iff its occurrence in $F$ is positive (resp. negative). Such a quasiatom $G$ is **elementary** iff it is an elementary atom; otherwise we say that it is **nonelementary**.

In Section 6 of [3] we defined the term "$E$-specification" ("$E$-specifies"). This terminology straightforwardly extends to hyperformulas and their quasiatoms. Note that a quasiatom of a given hyperformula $F$ is uniquely determined by its $F$-specification, even though this is generally not so for subhyperformulas that are not quasiatoms. For example, where $E = q \vee \neg p$, the string '2.' uniquely $E$-specifies the quasiatom $p$ while it $E$-specifies the occurences of two subformulas: $p$ and $\neg p$. Since specifications uniquely determine quasiatoms, we can use phrases such as "the occurrence $\gamma$ of $G$ in $E$" as long as $G$ is a quasiatom, meaning the occurrence of $G$ in $E$ that is $E$-specified by string $\gamma$.

For a run $\Gamma$, hyperformula $F$ and quasiatom $H$ of $F$ that is $F$-specified by $\gamma$, we use the notation $\Gamma^\gamma_F$ defined by:

$$\Gamma^\gamma_F = \begin{cases} \Gamma^\gamma & \text{if } H \text{ is positive in } F; \\ \neg \Gamma^\gamma & \text{if } H \text{ is negative in } F. \end{cases}$$

By the **surface complexity** of a hyperformula $F$ we mean the number of surface occurrences of $\neg$, $\wedge$, $\vee$, $\rightarrow$ in $F$. Some proofs in this paper will employ induction on surface complexity.

**Lemma 3.7** *Assume $^*$ is a perfect interpretation, $E$ is any hyperformula, and $\Gamma$ is any run. Then $\Gamma \in \mathbf{LR}^{E^*}$ iff every labeled move of $\Gamma$ has the form $\wp\gamma\beta$ for some $\gamma$ that $E$-specifies a nonelementary quasiatom $F$ of $E$, such that $\Gamma^\gamma_E \in \mathbf{LR}^{F^*}$.*

**Proof.** Assume the conditions of the lemma. Since $^*$ is perfect, all the formulas that we deal with are interpreted as constant games and, as noted earlier, we do not need to bother about the distinction between $\mathbf{Lr}$ and $\mathbf{LR}$. We prove this lemma by induction on the surface complexity of $E$. In doing so, we can safely assume that $E$ is just a $(\neg, \vee)$-combination of quasiatoms. Indeed, if not, every subhyperformula $K \rightarrow L$ of $E$ can be rewritten as $\neg K \vee L$, and every subhyperformula $K_1 \wedge \ldots \wedge K_n$ rewritten as $\neg(\neg K_1 \vee \ldots \vee \neg K_n)$. It is not hard to see that each quasiatom of the resulting hyperformula $E'$ will be $E'$-specified by the same string as the string that $E$-specifies that quasiatom, and the positive/negative status of quasiatoms will also be identical in $E$ and $E'$; next, by Definition 3.2(8) and Exercise 3.3(2) of [3], we will have $E^* = E'^*$, so that $E$ and $E'$ will be the "same" in every relevant aspect.

For the basis of induction, assume $E$ is a quasiatom. If $E$ is an elementary quasiatom and hence $E^*$ is an elementary game, then $\Gamma \in \mathbf{LR}^{E^*}$ iff $\Gamma = \langle \rangle$, because, as we remember, $\langle \rangle$ is the only legal run of elementary games. And, since in this case $E$ has no nonelementary quasiatoms, the condition $\Gamma = \langle \rangle$ is, in turn, equivalent to the condition "every labeled move of $\Gamma$ has the form $\wp\gamma\beta$ for some $\gamma$ that $E$-specifies a nonelementary quasiatom $F$ of $E$ ...". Thus, everything is as claimed by the lemma. Suppose now $E$ is a nonelementary quasiatom. Note that the occurrence of $E$ — the only nonelementary quasiatom $F = E$ of $E$ — in itself is $E$-specified by the empty string $\epsilon$. Inserting $\epsilon$ does not change a string, so every labeled move $\wp\beta$ of $\Gamma$ has the form $\wp\epsilon\beta$. And, of course, $\Gamma^\epsilon_E = \Gamma$. In view of these observations, the claim of the lemma is trivially satisfied.

For the inductive step, assume $E = \neg K$. By the definition of $\neg$, $\Gamma \in \mathbf{LR}^{E^*}$ iff $\neg\Gamma \in \mathbf{LR}^{K^*}$. In turn, by the induction hypothesis, $\neg\Gamma \in \mathbf{LR}^{K^*}$ iff every labeled move of $\neg\Gamma$ has the form $\wp\gamma\beta$ for some $\gamma$ that $K$-specifies a nonelementary quasiatom $F$ of $K$, such that $(\neg\Gamma)^\gamma_K \in \mathbf{LR}^{F^*}$. But notice that the same $\gamma$ also $E$-specifies the same quasiatom $F$, and that $(\neg\Gamma)^\gamma_K = \Gamma^\gamma_E$. Hence, the statement of the lemma is correct.

Finally, assume $E = K_1 \vee \ldots \vee K_n$. By the definition of $\vee$, $\Gamma \in \mathbf{LR}^{E^*}$ iff every move of $\Gamma$ starts with '$i$.' for some $i \in \{1, \ldots, n\}$ and, for each such $i$, we have $\Gamma^{i.} \in \mathbf{LR}^{K_i^*}$. In turn, by the induction hypothesis, $\Gamma^{i.} \in \mathbf{LR}^{K_i^*}$ iff every labeled move of $\Gamma^{i.}$ has the form $\wp\delta\beta$ for some $\delta$ that $K_i$-specifies a nonelementary quasiatom $F$ of $K_i$, such that $(\Gamma^{i.})^\delta_{K_i} \in \mathbf{LR}^{F^*}$. Notice that the same $F$ is a nonelementary quasiatom of $E$ which is $E$-specified by $i.\delta$, and that $(\Gamma^{i.})^\delta_{K_i} = \Gamma^{i.\delta}_E$. Thus, $\Gamma \in \mathbf{LR}^{E^*}$ iff every labeled move of $\Gamma$ has the form $\wp i.\delta\beta$, where $i.\delta$ is the $E$-specification of a nonelementary quasiatom $F$ of $E$, such that $\Gamma^{i.\delta}_E \in \mathbf{LR}^{F^*}$. In other words, with $i.\delta$ in the role of $\gamma$, the statement of the lemma holds. $\square$

We will be using the expression

$$\Gamma^{-\gamma}$$

to denote the result of deleting in run $\Gamma$ every labeled move that is $\wp\gamma\beta$ for some player $\wp$ and move/string $\beta$. Do not be misled by the symmetry in notation: $\Gamma^{-\gamma}$ and $\Gamma^\gamma$ are not "dual" in any reasonable sense. Also, to avoid possible ambiguity, our present notational convention assumes that the symbol "$-$" never occurs in moves (otherwise replace it with a symbol that satisfies such a condition).

**Lemma 3.8** *Assume $*$ is a perfect interpretation, $E$ is any hyperformula, $\Phi$ is a legal position of $E^*$, $\gamma$ is the $E$-specification of a nonelementary quasiatom $F$ of $E$, $G$ is a hyperformula with $\langle\Phi_E^\gamma\rangle F^* = G^*$, and $H$ is the result of replacing $F$ by $G$ in $E$. Then $\langle\Phi\rangle E^* = \langle\Phi^{-\gamma}\rangle H^*$.*

**Proof.** Assume the conditions of the lemma. Our proof proceeds by induction on the surface complexity of $E$. As in the previous lemma, considering only $\neg$ and $\vee$ in the inductive step would be sufficient.

Assume $E$ is a quasiatom, so that $F = E$, $H = G$, and $\gamma$ is the empty string $\epsilon$. According to one of the assumptions of the lemma, $\langle\Phi_E^\epsilon\rangle F^* = G^*$. Hence, as $\Phi_E^\epsilon = \Phi$, we have $\langle\Phi\rangle F^* = G^*$. The equations $F = E$ and $H = G$ allow us to rewrite $\langle\Phi\rangle F^* = G^*$ as $\langle\Phi\rangle E^* = H^*$. Of course $H^* = \langle\rangle H^*$, and thus $\langle\Phi\rangle E^* = \langle\rangle H^*$. But notice that $\langle\rangle = \langle\Phi^{-\epsilon}\rangle$, which yields the desired $\langle\Phi\rangle E^* = \langle\Phi^{-\epsilon}\rangle H^*$.

Next, assume $E = \neg K$. As in the corresponding step of the previous lemma, $\gamma$ remains the $K$-specification of $F$, and $\Phi_E^\gamma = (\neg\Phi)_K^\gamma$. Also, $\neg\Phi \in \mathbf{LR}^{K^*}$. It is our assumption that $\langle\Phi_E^\gamma\rangle F^* = G^*$, and therefore $\langle(\neg\Phi)_K^\gamma\rangle F^* = G^*$. Then, by the induction hypothesis, $\langle\neg\Phi\rangle K^* = \langle(\neg\Phi)^{-\gamma}\rangle L^*$, where $L$ is the result of replacing $F$ by $G$ in $K$. Hence $\neg(\langle\neg\Phi\rangle K^*) = \neg(\langle(\neg\Phi)^{-\gamma}\rangle L^*)$. By Lemma 3.5, $\neg(\langle\neg\Phi\rangle K^*) = \langle\Phi\rangle E^*$ and $\neg(\langle(\neg\Phi)^{-\gamma}\rangle L^*) = \langle\Phi^{-\gamma}\rangle\neg L^*$. Consequently, $\langle\Phi\rangle E^* = \langle\Phi^{-\gamma}\rangle\neg L^*$. But, of course, $\neg L^* = H^*$. Thus, $\langle\Phi\rangle E^* = \langle\Phi^{-\gamma}\rangle H^*$.

Finally, assume

$$E = K_1 \vee K_2 \vee \ldots \vee K_n.$$

Then, by Lemma 3.6,

$$\langle\Phi\rangle E^* = \langle\Phi^{1.}\rangle K_1^* \vee \langle\Phi^{2.}\rangle K_2^* \vee \ldots \vee \langle\Phi^{n.}\rangle K_n^*. \tag{6}$$

Let $K_i$ be the disjunct of $E$ that contains $F$, and let $\delta$ be the $K_i$-specification of $F$. For simplicity of representation and without loss of generality, let us assume here that $i = 1$ — of course, any other $i$ can be handled in a similar way. Thus, $\gamma = 1.\delta$, and we have

$$H = L \vee K_2 \vee \ldots \vee K_n, \tag{7}$$

where $L$ is the result of replacing $F$ by $G$ in $K_1$.

As $\gamma = 1.\delta$, for any $j \neq 1$ we obviously have $\Phi^{j.} = (\Phi^{-\gamma})^{j.}$. Hence, (6) can be rewritten as

$$\langle\Phi\rangle E^* = \langle\Phi^{1.}\rangle K_1^* \vee \langle(\Phi^{-\gamma})^{2.}\rangle K_2^* \vee \ldots \vee \langle(\Phi^{-\gamma})^{n.}\rangle K_n^*. \tag{8}$$

It is our assumption that $\langle\Phi_E^\gamma\rangle F^* = G^*$, i.e. $\langle\Phi_E^{1.\delta}\rangle F^* = G^*$. But obviously $\Phi_E^{1.\delta} = (\Phi^{1.})_{K_1}^\delta$, and therefore $\langle(\Phi^{1.})_{K_1}^\delta\rangle F^* = G^*$. Then, by the induction hypothesis, $\langle\Phi^{1.}\rangle K_1^* = \langle(\Phi^{1.})^{-\delta}\rangle L^*$. But it is not hard to see that $(\Phi^{1.})^{-\delta} = (\Phi^{-1.\delta})^{1.}$. Hence, $\langle\Phi^{1.}\rangle K_1^* = \langle(\Phi^{-1.\delta})^{1.}\rangle L^*$, i.e. $\langle\Phi^{1.}\rangle K_1^* = \langle(\Phi^{-\gamma})^{1.}\rangle L^*$. This allows us to rewrite (8) as

$$\langle\Phi\rangle E^* = \langle(\Phi^{-\gamma})^{1.}\rangle L^* \vee \langle(\Phi^{-\gamma})^{2.}\rangle K_2^* \vee \ldots \vee \langle(\Phi^{-\gamma})^{n.}\rangle K_n^*. \tag{9}$$

Since $\Phi \in \mathbf{LR}^{E^*}$, every move of ($\Phi$ and hence of) $\Phi^{-\gamma}$ starts with '$i.$' for some $i \in \{1,\ldots,n\}$. And, with Convention 3.3 in mind, (9) implies that

$$(\Phi^{-\gamma})^{1.} \in \mathbf{LR}^{L^*},\ (\Phi^{-\gamma})^{2.} \in \mathbf{LR}^{K_2^*},\ \ldots\ (\Phi^{-\gamma})^{2.} \in \mathbf{LR}^{K_2^*}.$$

By the definition of $\vee$, all this means that $\Phi^{-\gamma} \in \mathbf{LR}^{L^* \vee K_2^* \vee \ldots \vee K_n^*}$. Then, by Lemma 3.6,

$$\langle\Phi^{-\gamma}\rangle(L^* \vee K_2^* \vee \ldots \vee K_n^*) = \langle(\Phi^{-\gamma})^{1.}\rangle L^* \vee \langle(\Phi^{-\gamma})^{2.}\rangle K_2^* \vee \ldots \vee \langle(\Phi^{-\gamma})^{n.}\rangle K_n^*.$$

The above, in conjunction with (9) and (7), yields the desired $\langle\Phi\rangle E^* = \langle\Phi^{-\gamma}\rangle H^*$. □

## 3.4 Manageability

**Definition 3.9** Let $F$ be a balanced hyperformula. We say that a run $\Gamma$ is $F$**-manageable** iff the following three conditions are satisfied:

1. Every labeled move of $\Gamma$ has the form $\wp\gamma\alpha$, where $\gamma$ is the $F$-specification of a surface occurrence of either a general atom, or a hybrid atom.

2. If $\gamma$ is the $F$-specification of a general atom, then $\Gamma^\gamma$ entirely consists of $\bot$-labeled moves.

3. If, for some hybrid atom $P_q$, $\pi$ and $\nu$ are the $F$-specifications of the positive and the negative occurrence of $P_q$ in $F$, respectively, then $\Gamma^\pi$ is a $\top$-delay (see [3], Section 3) of $\neg\Gamma^\nu$.

The above technical concept will play a central role in our soundness proof for **CL2**. Very roughly, the intuition here is that, when $\Gamma$ is $F$-manageable, playing it in no way affects the logical structure of $F$ (clause 1), ensures that the subgames in the "matched" occurrences of atoms evolve to — in a sense — the same games (clause 3), and that $\top$ does not make any hasty moves in unmatched atoms (clause 2), so that, if and when at some later point such an atom finds a match, $\top$ will still have a chance to "even out" the corresponding two subgames.

**Lemma 3.10** *Let $E$ be any balanced hyperformula, $*$ any perfect interpretation, and $\Gamma$ an infinite run with arbitrarily long finite initial segments that are $E$-manageable legal positions of $E^*$. Then $\Gamma$ is an $E$-manageable legal run of $E^*$.*

**Proof.** Assume the conditions of the lemma. They imply that every finite initial segment of $\Gamma$ is a legal position of $E^*$. Hence, by condition (a) of Definition 3.1 of [3], $\Gamma$ is a legal run of $E^*$. Also, obviously $\Gamma$ satisfies conditions 1 and 2 of Definition 3.9 because it has arbitrarily long initial segments that satisfy those conditions. So, what remains to show is that $\Gamma$ satisfies condition 3 of Definition 3.9.

Suppose, for a contradiction, that this is not the case. In particular, there are $\pi$, $\nu$ and $P_q$ as described in the antecedent of condition 3 such that $\Gamma^\pi$ is not a $\top$-delay of $\neg\Gamma^\nu$. This means that at least one of the following two statements is true:

**(i)** For one of the players $\wp$, the subsequence of the $\wp$-labeled moves of $\Gamma^\pi$ (i.e. the result of deleting in $\Gamma^\pi$ all $\neg\wp$-labeled moves) is not the same as that of $\neg\Gamma^\nu$, or

**(ii)** For some $k, n$, in $\neg\Gamma^\nu$ the $n$th $\top$-labeled move is made later than the $k$th $\bot$-labeled move, but in $\Gamma^\pi$ the $n$th $\top$-labeled move is made earlier than the $k$th $\bot$-labeled move.

Whether (i) or (ii) is the case, it is not hard to see that, beginning from some (finite) $n$, every initial segment $\Psi$ of $\Gamma$ of length $\geq n$ will satisfy the same (i) or (ii) in the role of $\Gamma$, and hence $\Psi$ will not be an $E$-manageable position of $E^*$. This contradicts the assumptions of our lemma. $\square$

**Lemma 3.11** *Assume $E$ is a balanced hyperformula, $*$ is a perfect interpretation, and $\Omega$ is an $E$-manageable legal position of $E^*$. Suppose $\gamma$ is the $E$-specification of a negative (resp. positive) surface occurrence of a subformula $G_1 \sqcap \ldots \sqcap G_n$ (resp. $G_1 \sqcup \ldots \sqcup G_n$), and $i \in \{1, \ldots, n\}$. Let $H$ be the result of substituting in $E$ the above occurrence by $G_i$. Then:*

1. *$\langle\Omega\rangle$ is $H$-manageable;*

2. *$\langle\Omega, \top\gamma i\rangle E^* = \langle\Omega\rangle H^*$.*

**Proof.** Assume the conditions of the lemma. It is not hard to see that each of the three conditions of Definition 3.9 is inherited by $H$ from $E$. This implies clause 1. Next, since $\Omega$ does not contain $\gamma$-prefixed moves (for otherwise, by condition 1 of Definition 3.9, it would not be $E$-manageable), we have $\langle\Omega, \top\gamma i\rangle^\gamma = \langle\top i\rangle$ and $\langle\Omega, \top\gamma i\rangle^{\neg\gamma} = \Omega$. Based on this, clause 2 of the present lemma is an easy consequence of our Lemmas 3.7, 3.8 and clauses 5,6 of Lemma 3.7 of [3]. $\square$

**Lemma 3.12** *Assume $A$ is a constant static game, $\wp$ is either player, and $\Gamma, \Delta$ are runs such that $\Delta$ is a $\wp$-delay of $\Gamma$. Then:*

1. *If $\Delta$ is a $\wp$-illegal run of $A$, then so is $\Gamma$.*

2. *If $\Gamma$ is a $\neg\wp$-illegal run of $A$, then so is $\Delta$.*

**Proof.** The above is a fact known from [2] (Lemma 4.7). □

**Lemma 3.13** *Assume $E$ is a balanced hyperformula, $*$ is a perfect interpretation, and $\Omega$ is an $E$-manageable legal position of $E^*$. Suppose $H$ is the hyperformula that results from $E$ by replacing in it a positive surface occurrence $\pi$ and a negative surface occurrence $\nu$ of a general atom $P$ by a hybrid atom $P_q$, such that $H$ remains balanced. Further assume that $\Omega^\pi = \langle \bot\pi_1, \ldots, \bot\pi_n \rangle$ and $\Omega^\nu = \langle \bot\nu_1, \ldots, \bot\nu_m \rangle$. Then $\langle \Omega, \top\pi\nu_1, \ldots, \top\pi\nu_m, \top\nu\pi_1, \ldots, \top\nu\pi_n \rangle$ is an $H$-manageable legal position of $H^*$.*

**Proof.** Assume the conditions of the lemma. Let
$$\Phi = \langle \Omega, \top\pi\nu_1, \ldots, \top\pi\nu_m, \top\nu\pi_1, \ldots, \top\nu\pi_n \rangle.$$

Notice that
$$\Phi_H^\pi = \Phi^\pi = \langle \bot\pi_1, \ldots, \bot\pi_n, \top\nu_1, \ldots, \top\nu_m \rangle; \tag{10}$$
$$\Phi_H^\nu = \neg\Phi^\nu = \langle \top\nu_1, \ldots, \top\nu_m, \bot\pi_1, \ldots, \bot\pi_n \rangle. \tag{11}$$

Thus,
$$\Phi_H^\pi \text{ is a } \top\text{-delay of } \Phi_H^\nu. \tag{12}$$

This implies that $\Phi$ is $H$-manageable, because conditions 1 and 2 of Definition 3.9 are obviously inherited by $H, \Phi$ from $E, \Omega$, and so is condition 3 for any hybrid atom $R_s$ different from $P_q$.

What remains to show is that $\Phi \in \mathbf{LR}^{H^*}$. For this, in view of Lemma 3.7, it would be sufficient to verify that $\Phi_H^\pi \in \mathbf{LR}^{P^*}$ and $\Phi_H^\nu \in \mathbf{LR}^{P^*}$, because for any other (different from $\pi, \nu$) relevant $\gamma$, the similar condition is inherited by $H, \Phi$ from $E, \Omega$ taking into account that $\Phi^\gamma = \Omega^\gamma$.

Since $\Omega$ is a legal position of $E^*$, by Lemma 3.7, we have both $\Omega_E^\pi \in \mathbf{LR}^{P^*}$ and $\Omega_E^\nu \in \mathbf{LR}^{P^*}$. Thus,
$$\langle \bot\pi_1, \ldots, \bot\pi_n \rangle \in \mathbf{LR}^{P^*}; \tag{13}$$
$$\langle \top\nu_1, \ldots, \top\nu_m \rangle \in \mathbf{LR}^{P^*}. \tag{14}$$

Suppose $\Phi_H^\pi \notin \mathbf{LR}^{P^*}$, i.e. (by (10)) $\langle \bot\pi_1, \ldots, \bot\pi_n, \top\nu_1, \ldots, \top\nu_m \rangle \notin \mathbf{LR}^{P^*}$. In view of (13), $\Phi_H^\pi$ cannot be a $\bot$-illegal position of $P^*$. So, it must be $\top$-illegal. But then, by (12) and clause 1 of Lemma 3.12, $\Phi_H^\nu$ is a $\top$-illegal position of $P^*$. This, however, is in obvious contradiction with (11) and (14).

Suppose now $\Phi_H^\nu \notin \mathbf{LR}^{P^*}$, i.e. (by (11)) $\langle \top\nu_1, \ldots, \top\nu_m, \bot\pi_1, \ldots, \bot\pi_n \rangle \notin \mathbf{LR}^{P^*}$. This case is similar/symmetric to the previous one. In view of (14), $\Phi_H^\nu$ cannot be a $\top$-illegal position of $P^*$. So, it must be $\bot$-illegal. But then, by (12) and clause 2 of Lemma 3.12, $\Phi_H^\pi$ is a $\bot$-illegal position of $P^*$. This, however, is in contradiction with (10) and (13). □

**Lemma 3.14** *Assume $E$ is a balanced hyperformula, $\alpha$ is any move, $*$ is a perfect interpretation, $\Omega$ is an $E$-manageable position, and $\langle \Omega, \bot\alpha \rangle \in \mathbf{LR}^{E^*}$. Then one of the following conditions is satisfied:*

1. *$\alpha = \gamma\beta$, where $\gamma$ is the $E$-specification of a surface occurrence of a general atom in $E$. In this case $\langle \Omega, \bot\alpha \rangle$ is an $E$-manageable legal position of $E^*$.*

2. *$\alpha = \gamma\beta$, where $\gamma$ is the $E$-specification of a surface occurrence of a hybrid atom in $E$. Let $\sigma$ be the $E$-specification of the other occurrence of the same hybrid atom. Then $\langle \Omega, \bot\gamma\beta, \top\sigma\beta \rangle$ is an $E$-manageable legal position of $E^*$.*

3. *$\alpha = \gamma i$, where $\gamma$ is the $E$-specification of a positive (resp. negative) surface occurrence of a subformula $G_1 \sqcap \ldots \sqcap G_n$ (resp. $G_1 \sqcup \ldots \sqcup G_n$) and $i \in \{1, \ldots, n\}$. In this case, where $H$ is the result of substituting in $E$ the above occurrence by $G_i$, we have:*

   (a) *$\langle \Omega \rangle$ is $H$-manageable;*
   
   (b) *$\langle \Omega, \bot\alpha \rangle E^* = \langle \Omega \rangle H^*$.*

**Proof.** Let $E$, $\alpha$, $*$, $\Omega$ be as assumed in the lemma. By Lemma 3.7, the condition $\langle \Omega, \bot\alpha \rangle \in \mathbf{Lr}^{E^*}$ implies that $\alpha$ should be $\gamma\beta$, where $\gamma$ is the $E$-specification of a nonelementary quasiatom $F$ of $E$. Fix these $\gamma$, $\beta$ and $F$. By the definition of nonelementary quasiatom, $F$ is either (1) a general atom, or (2) a hybrid atom, or (3) a (hyper)formula of the form $G_1 \sqcap \ldots \sqcap G_n$ or $G_1 \sqcup \ldots \sqcup G_n$. We consider each of these three possibilities separately.

*Case 1:* $F$ is a general atom. Obviously adding to a $E$-manageable position ($\Omega$) a $\bot$-labeled move whose prefix $E$-specifies a surface occurrence of a general atom again yields an $E$-manageable position. So, $\langle \Omega, \bot\alpha \rangle$ is $E$-manageable; by the assumptions of the lemma, it is also a legal position of $E^*$. Thus, condition 1 of Lemma 3.14 is satisfied.

*Case 2:* $F$ is a hybrid atom $P_q$. Let $\sigma$ be the $E$-specification of the other occurrence of $P_q$. We want to show that $\langle \Omega, \bot\gamma\beta, \top\sigma\beta \rangle$ is an $E$-manageable legal position of $E^*$ and hence condition 2 of the lemma is satisfied.

*Subcase 2.1.* Assume $F$ is negative in $E$. That is, $\gamma$ $E$-specifies the negative occurrence of $P_q$ and $\sigma$ $E$-specifies the positive occurrence of $P_q$. Since $\Omega$ is $E$-manageable, $\Omega^\sigma$ is a $\top$-delay of $\neg\Omega^\gamma$. Obviously this implies that $\langle \Omega^\sigma, \top\beta \rangle$ is a $\top$-delay of $\langle \neg\Omega^\gamma, \top\beta \rangle = \neg\langle \Omega^\gamma, \bot\beta \rangle$. From here, observing that $\langle \Omega^\sigma, \top\beta \rangle = \langle \Omega, \bot\gamma\beta, \top\sigma\beta \rangle^\sigma$ and $\langle \Omega^\gamma, \bot\beta \rangle = \langle \Omega, \bot\gamma\beta, \top\sigma\beta \rangle^\gamma$, we get:

$$\langle \Omega, \bot\gamma\beta, \top\sigma\beta \rangle^\sigma \text{ is a } \top\text{-delay of } \neg\langle \Omega, \bot\gamma\beta, \top\sigma\beta \rangle^\gamma. \tag{15}$$

Remembering our assumption that $\Omega$ is $E$-manageable and taking into account that for any $\delta \neq \gamma, \sigma$ we have $\langle \Omega, \bot\gamma\beta, \top\sigma\beta \rangle^\delta = \Omega^\delta$, (15) is obviously sufficient to conclude that

$$\langle \Omega, \bot\gamma\beta, \top\sigma\beta \rangle \text{ is } E\text{-manageable.} \tag{16}$$

According to the assumptions of the lemma, $\langle \Omega, \bot\gamma\beta \rangle \in \mathbf{LR}^{E^*}$. By Lemma 3.7, this implies that $\langle \Omega, \bot\gamma\beta \rangle_E^\gamma \in \mathbf{LR}^{P^*}$, i.e. $\neg\langle \Omega, \bot\gamma\beta \rangle^\gamma \in \mathbf{LR}^{P^*}$ and, as $\gamma \neq \sigma$, we clearly have $\neg\langle \Omega, \bot\gamma\beta, \top\sigma\beta \rangle^\gamma \in \mathbf{LR}^{P^*}$. Then, by (15) and clause 1 of Lemma 3.12, $\langle \Omega, \bot\gamma\beta, \top\sigma\beta \rangle^\sigma$ is not a $\top$-illegal position of $P^*$. It is not $\bot$-illegal either, for otherwise we would have $\langle \Omega, \bot\gamma\beta \rangle^\sigma \notin \mathbf{LR}^{P^*}$ which, in view of Lemma 3.7, contradicts our assumption that $\langle \Omega, \bot\gamma\beta \rangle \in \mathbf{LR}^{E^*}$. Thus, $\langle \Omega, \bot\gamma\beta, \top\sigma\beta \rangle^\sigma \in \mathbf{LR}^{P^*}$. Taking into account that for any $\delta \neq \sigma$ we have $\langle \Omega, \bot\gamma\beta, \top\sigma\beta \rangle^\delta = \langle \Omega, \bot\gamma\beta \rangle^\delta$, Lemma 3.7 in conjunction with our assumption $\langle \Omega, \bot\gamma\beta \rangle \in \mathbf{LR}^{E^*}$ then implies that $\langle \Omega, \bot\gamma\beta, \top\sigma\beta \rangle \in \mathbf{LR}^{E^*}$. This, together with (16), means that condition 2 of Lemma 3.14 is satisfied.

*Subcase 2.2.* Now assume $F$ is positive in $E$. That is, $\gamma$ $E$-specifies the positive occurrence of $P_q$ and $\sigma$ $E$-specifies the negative occurrence of $P_q$. By the $E$-manageability of $\Omega$, $\Omega^\gamma$ is a $\top$-delay of $\neg\Omega^\sigma$. Hence $\langle \Omega^\gamma, \bot\beta \rangle$ is a $\top$-delay of $\langle \neg\Omega^\sigma, \bot\beta \rangle = \neg\langle \Omega^\sigma, \top\beta \rangle$. Then, as $\langle \Omega^\gamma, \bot\beta \rangle = \langle \Omega, \bot\gamma\beta, \top\sigma\beta \rangle^\gamma$ and $\langle \Omega^\sigma, \top\beta \rangle = \langle \Omega, \bot\gamma\beta, \top\sigma\beta \rangle^\sigma$, we get:

$$\langle \Omega, \bot\gamma\beta, \top\sigma\beta \rangle^\gamma \text{ is a } \top\text{-delay of } \neg\langle \Omega, \bot\gamma\beta, \top\sigma\beta \rangle^\sigma. \tag{17}$$

From here, just as in Subcase 2.1, we conclude that statement (16) is true.

Next, as noted in Subcase 2.1, $\langle \Omega, \bot\gamma\beta \rangle_E^\gamma \in \mathbf{LR}^{P^*}$, which now simply means that $\langle \Omega, \bot\gamma\beta \rangle^\gamma \in \mathbf{LR}^{P^*}$. Then $\langle \Omega, \bot\gamma\beta, \top\sigma\beta \rangle^\gamma$, which equals $\langle \Omega, \bot\gamma\beta \rangle^\gamma$, is not a $\bot$-illegal position of $P^*$. Therefore, by (17) and clause 2 of Lemma 3.12, $\neg\langle \Omega, \bot\gamma\beta, \top\sigma\beta \rangle^\sigma$ is not a $\bot$-illegal position of $P^*$. But obviously this position — let us rewrite it as $\langle \neg\Omega, \top\gamma\beta, \bot\sigma\beta \rangle^\sigma$ — is not $\top$-illegal either, for otherwise we would have $\langle \neg\Omega, \top\gamma\beta \rangle^\sigma \notin \mathbf{LR}^{P^*}$, i.e. $\neg\langle \Omega, \bot\gamma\beta \rangle^\sigma \notin \mathbf{LR}^{P^*}$, i.e. $\langle \Omega, \bot\gamma\beta \rangle_E^\sigma \notin \mathbf{LR}^{P^*}$ which, by Lemma 3.7, contradicts our assumption that $\langle \Omega, \bot\gamma\beta \rangle \in \mathbf{LR}^{E^*}$. Thus, $\neg\langle \Omega, \bot\gamma\beta, \top\sigma\beta \rangle^\sigma \in \mathbf{LR}^{P^*}$. Continuing as in Subcase 2.1, we can conclude that condition 2 of Lemma 3.14 is satisfied.

*Case 3:* $F$ is $G_1 \sqcap \ldots \sqcap G_n$ or $G_1 \sqcup \ldots \sqcup G_n$. We want to show that then (the rest of) condition 3 of Lemma 3.14 is satisfied. Since $\Omega$ is $E$-manageable, clause 1 of Definition 3.9 imples $\Omega$ does not contain $\gamma$-prefixed moves. Therefore we have:

$$\langle \Omega, \bot\gamma\beta \rangle^\gamma = \langle \bot\gamma\beta \rangle^\gamma; \tag{18}$$

$$\langle \Omega, \bot\gamma\beta \rangle^{\neg\gamma} = \Omega. \tag{19}$$

By (18) and Lemma 3.7, $\langle \bot\gamma\beta \rangle_E^\gamma \in \mathbf{LR}^{F^*}$. In view of clauses 5(a) and 6(a) of Lemma 3.7 of [3], this is the case when $\beta = i \in \{1, \ldots, n\}$ and either $F = G_1 \sqcap \ldots \sqcap G_n$ and $F$ is positive in $E$ (so that $\langle \bot\gamma\beta \rangle_E^\gamma = \langle \bot i \rangle$), or $F = G_1 \sqcup \ldots \sqcup G_n$ and $F$ is negative in $E$ (so that $\langle\bot\gamma\beta\rangle^\gamma_E = \langle\top i\rangle$). In either case, by clauses 5(b) and 6(b) of the same lemma, we have $\langle\bot\gamma\beta\rangle^\gamma_E F^* = G_i^*$. By (18), this means that $\langle\Omega, \bot\gamma\beta\rangle^\gamma_E F^* = G_i^*$. Then, according to Lemma 3.8, $\langle\Omega, \bot\gamma\beta\rangle E^* = \langle\Omega, \bot\gamma\beta\rangle^{-\gamma} H^*$, where $H$ is the result of replacing in $E$ the quasiatom $F$ by $G_i$. Applying (19) and changing $\gamma\beta$ back to $\alpha$, the just-derived equation can be rewritten as $\langle\Omega, \bot\alpha\rangle E^* = \langle\Omega\rangle H^*$. To conclude that condition 3 of Lemma 3.14 is satisfied, it remains to notice that $\Omega$ is $H$-manageable. This is so for the same reasons as in Lemma 3.11. $\square$

## 3.5 Finalization

Remember from [3] that we call the constant elementary games $\top$ and $\bot$ *trivial*. Note that when a formula $E$ is elementary and an interpretation $*$ is perfect, $E^*$ is a trivial game. Here we define the relation $\leq$ on trivial games by stipulating that $A \leq B$ iff $A = \bot$ or $B = \top$. In other words, $A \leq B$ iff $(A \to B) = \top$.

**Lemma 3.15** *Suppose $*$ is a perfect interpretation, $F_G$ is an elementary formula, and $F_H$ is the result of replacing in $F_G$ a quasiatom $G$ by an elementary formula $H$. Then we have $F_G^* \leq F_H^*$ as long as one of the following two conditions is satisfied:*

- *$G$ is positive and $G^* \leq H^*$;*
- *$G$ is negative and $H^* \leq G^*$.*

**Proof.** The above lemma does nothing but rephrases, in our terms, a known fact from classical logic according to which, if in an interpreted formula $F$ we replace a positive (resp. negative) occurrence of a subformula $G$ by a formula $H$ whose Boolean value is not less (resp. not greater) than that of $G$, then the value of the resulting formula will not be less than that of $F$. $\square$

Now we introduce the operation $\langle\Gamma\rangle\!\downarrow\! A$ of the type $\{runs\} \times \{constant\ games\} \to \{trivial\ games\}$, which is rather similar to prefixation. Intuitively $\langle\Gamma\rangle\!\downarrow\! A$, that we call the $\Gamma$**-finalization** of $A$, is the proposition "$\Gamma$ is a $\top$-won run of $A$". This operation is only defined when $\Gamma$ is a legal run of $A$. Extending Convention 3.3 to finalization, every time we make a statement that applies $\langle\Gamma\rangle\!\downarrow$ to a constant game $A$, we imply that $\Gamma$ is a legal run of $A$ and hence $\langle\Gamma\rangle\!\downarrow\! A$ is defined. Here is the formal definition of the operation of finalization:

**Definition 3.16** Assume $A$ is a constant game and $\Gamma$ a legal run of $A$. Then $\langle\Gamma\rangle\!\downarrow\! A$ is the trivial game defined by $\mathbf{Wn}^{\langle\Gamma\rangle\downarrow A}\langle\rangle = \mathbf{Wn}^A\langle\Gamma\rangle$.

**Lemma 3.17** *For any constant game $A$ and run $\Gamma$ with $\Gamma \in \mathbf{LR}^{\neg A}$ — or, equivalently, $\neg\Gamma \in \mathbf{LR}^A$ — we have $\langle\Gamma\rangle\!\downarrow\!\neg A = \neg(\langle\neg\Gamma\rangle\!\downarrow\! A)$.*

**Proof.** It is safe to identify players $\top, \bot$ with the corresponding two trivial games $\top, \bot$. In particular, for a constant game $B$, we can use $\langle\Gamma\rangle\!\downarrow\! B$ and $\mathbf{Wn}^B\langle\Gamma\rangle$ interchangeably, even though, formally the former is a (trivial) game while the latter is a player.

Assume $A$ is a constant game, and $\Gamma \in \mathbf{LR}^{\neg A}$ which, by definition, means the same as $\neg\Gamma \in \mathbf{LR}^A$. By the definition of finalization, $\langle\Gamma\rangle\!\downarrow\!\neg A = \mathbf{Wn}^{\neg A}\langle\Gamma\rangle$; by the definition of $\neg$, $\mathbf{Wn}^{\neg A}\langle\Gamma\rangle = \neg\mathbf{Wn}^A\langle\neg\Gamma\rangle$; again by the definition of finalization, $\mathbf{Wn}^A\langle\neg\Gamma\rangle = \langle\neg\Gamma\rangle\!\downarrow\! A$ and hence $\neg\mathbf{Wn}^A\langle\neg\Gamma\rangle = \neg(\langle\neg\Gamma\rangle\!\downarrow\! A)$. These equations yield the desired $\langle\Gamma\rangle\!\downarrow\!(\neg A) = \neg(\langle\neg\Gamma\rangle\!\downarrow\! A)$. $\square$

**Lemma 3.18** *For any constant games $A_1, \ldots, A_n$ ($n \geq 2$) and run $\Gamma$ with $\Gamma \in \mathbf{LR}^{A_1 \vee \ldots \vee A_n}$, we have*

$$\langle\Gamma\rangle\!\downarrow\!(A_1 \vee \ldots \vee A_n) = \langle\Gamma^{1.}\rangle\!\downarrow\! A_1 \vee \ldots \vee \langle\Gamma^{n.}\rangle\!\downarrow\! A_n.$$

**Proof.** As in the previous lemma, we identify players $\top, \bot$ with the corresponding trivial games $\top, \bot$. Assume $A_1, \ldots, A_n$ are constant games and $\Gamma \in \mathbf{LR}^{A_1 \vee \ldots \vee A_n}$. The definition of $\vee$ guarantees that each $\Gamma^{i.}$ is in $\mathbf{LR}^{A_i}$ and hence the $\langle\Gamma^{i.}\rangle\!\downarrow\! A_i$ are defined. By the definition of finalization, $\langle\Gamma\rangle\!\downarrow\!(A_1 \vee \ldots \vee A_n) = \mathbf{Wn}^{A_1 \vee \ldots \vee A_n}\langle\Gamma\rangle$. The definition of $\vee$ easily implies that $\mathbf{Wn}^{A_1 \vee \ldots \vee A_n}\langle\Gamma\rangle = \mathbf{Wn}^{A_1}\langle\Gamma^{1.}\rangle \vee \ldots \vee \mathbf{Wn}^{A_n}\langle\Gamma^{n.}\rangle$. In turn, again by the definition of finalization, each $\mathbf{Wn}^{A_i}\langle\Gamma^{i.}\rangle$ is nothing but $\langle\Gamma^{i.}\rangle\!\downarrow\! A_i$. Putting all this together yields the desired $\langle\Gamma\rangle\!\downarrow\!(A_1 \vee \ldots \vee A_n) = (\langle\Gamma^{1.}\rangle\!\downarrow\! A_1) \vee \ldots \vee (\langle\Gamma^{n.}\rangle\!\downarrow\! A_n)$. $\square$

**Lemma 3.19** *For any constant games $A_1, \ldots, A_n$ ($n \geq 2$), we have:*

1. $\langle\rangle\downarrow(A_1 \sqcap \ldots \sqcap A_n) = \top$.

2. $\langle\rangle\downarrow(A_1 \sqcup \ldots \sqcup A_n) = \bot$.

**Proof.** Immediately from the relevant definitions. □

**Lemma 3.20** *Assume $*$ is a perfect interpretation, $E$ is any hyperformula, $\Gamma$ is a legal run of $E^*$, $\gamma$ is the $E$-specification of a nonelementary quasiatom $F$ of $E$, $G$ is an elementary formula with $\langle\Gamma_E^\gamma\rangle\downarrow F^* = G^*$, and $H$ is the result of replacing $F$ by $G$ in $E$. Then $\langle\Gamma\rangle\downarrow E^* = \langle\Gamma^{-\gamma}\rangle\downarrow H^*$.*

**Proof.** This lemma is very similar to Lemma 3.8, and the proof of the latter can be literally repeated here as long as in it we change $\Phi$ to $\Gamma$, replace prefixation by finalization, and replace the references to Lemmas 3.5 and 3.6 by references to Lemmas 3.17 and 3.18, respectively. □

**Lemma 3.21** *Assume $E$ is a stable balanced hyperformula, $*$ is a perfect interpretation, and $\Gamma$ is an $E$-manageable legal run of $E^*$. Then $\mathbf{Wn}^{E^*}\langle\Gamma\rangle = \top$.*

**Proof.** Assume the conditions of the lemma. For each $\gamma$ that $E$-specifies an occurrence of a general or hybrid atom, let us fix an elementary nonlogical atom $r_\gamma$ that does not occur in $E$ (neither directly nor as the elementary component of a hybrid atom). Since these atoms do not occur in $E$, we may make an arbitrary assumption regarding how they are interpreted by $*$ (otherwise replace $*$ by an appropriate interpretation). In particular, we assume that, for every $\gamma$ that $E$-specifies a quasiatom that is either a general atom $P$ or a hybrid atom $P_q$, $r_\gamma^* = \langle\Gamma_E^\gamma\rangle\downarrow P^*$. That each such $\langle\Gamma_E^\gamma\rangle\downarrow P^*$ is defined, i.e. $\Gamma_E^\gamma$ is in $\mathbf{LR}^{P^*}$, is guaranteed by Lemma 3.7.

Let $E_1$ denote the result of replacing in $E$:

- every surface occurrence of a general or hybrid atom by $r_\gamma$, where $\gamma$ is the $E$-specification of that occurrence;

- every surface occurrence of a subformula of the form $H_1 \sqcap \ldots \sqcap H_n$ by $\top$;

- every surface occurrence of a subformula of the form $H_1 \sqcup \ldots \sqcup H_n$ by $\bot$.

Since $\Gamma \in \mathbf{LR}^{E^*}$, by Lemma 3.7, every labeled move of $\Gamma$ has the form $\wp\gamma\beta$, where $\gamma$ is the $E$-specification of a nonelementary quasiatom $F$. If such an $F$ is $H_1 \sqcap \ldots \sqcap H_n$, as $\Gamma$ is $E$-manageable, we have $\Gamma_E^\gamma = \langle\rangle$ and hence, in view of clause 1 of Lemma 3.19, $\langle\Gamma_E^\gamma\rangle\downarrow F^* = \top$. Similarly, if $F$ is $H_1 \sqcup \ldots \sqcup H_n$, clause 2 of Lemma 3.19 yields $\langle\Gamma_E^\gamma\rangle\downarrow F^* = \bot$. Finally, if $F$ is a general or a hybrid atom, then, by our assumptions regarding how $*$ interprets $r_\gamma$, we have $\langle\Gamma_E^\gamma\rangle\downarrow F^* = r_\gamma^*$. In view of these observations, applying Lemma 3.20 as many times as the number of nonelementary quasiatoms of $E$, we get $\langle\Gamma\rangle\downarrow E^* = \langle\rangle\downarrow E_1^*$. But $E_1^*$ is an elementary game, and obviously for every elementary game $A$ we have $\langle\rangle\downarrow A = A$. Hence,

$$\langle\Gamma\rangle\downarrow E^* = E_1^*. \tag{20}$$

Assume $E$ contains $k$ hybrid atoms, where $q_1, \ldots, q_k$ are the elementary components of those atoms and $P_1, \ldots, P_k$ are the corresponding general components. Let $\pi_1, \ldots, \pi_k$ be the $E$-specifications of the positive occurrences of the hybrid atoms whose elementary components are $q_1, \ldots, q_k$, respectively. Similarly, let $\nu_1, \ldots, \nu_k$ be the $E$-specifications of the negative occurrences of the hybrid atoms whose elementary components are $q_1, \ldots, q_k$, respectively. Let $E_2$ be the result of replacing in $E_1$ each atom $r_{\pi_i}$ ($1 \leq i \leq k$) by $r_{\nu_i}$. According to clause 3 of Definition 3.9, for each such $1 \leq i \leq k$, $\Gamma^{\pi_i}$ is a $\top$-delay of $\neg\Gamma^{\nu_i}$. Hence, as $P_i^*$ is a static game, $\langle\neg\Gamma^{\nu_i}\rangle\downarrow P_i^* \leq \langle\Gamma^{\pi_i}\rangle\downarrow P_i^*$. This means that $r_{\nu_i}^* \leq r_{\pi_i}^*$. Then, applying Lemma 3.15 $k$ times, we get

$$E_2^* \leq E_1^*. \tag{21}$$

Next, assume $E$ contains $m$ negative occurrences of general atoms, $E$-specified by $\delta_1, \ldots, \delta_m$, and $n$ positive occurrences of general atoms, $E$-specified by $\sigma_1, \ldots, \sigma_n$. Let $E_3$ be the result of replacing in $E_2$ each atom $r_{\delta_i}$ ($1 \leq i \leq m$) by $\top$, and each atom $r_{\sigma_j}$ ($1 \leq j \leq n$) by $\bot$. Applying Lemma 3.15 $m+n$ times, we get
$$E_3^* \leq E_2^*. \tag{22}$$

Now compare $E_3$ with $\|E\|$. An analysis of how these two formulas have been obtained from $E$ can reveal that $E_3$ is just the result of replacing in $\|E\|$ all (both) occurrences of each atom $q_i$ from the earlier-discussed list $q_1,\ldots,q_k$ by $r_{\nu_i}$. That is, $E_3$ is a substitutional instance of $\|E\|$. The latter is classically valid because, by our assumptions, $E$ is stable. Therefore $E_3$ is also classically valid, and hence $E_3^* = \top$. Then statements (22), (21) and (20) yield $\langle\Gamma\rangle\downarrow E^* = \top$. This means nothing but that $\mathbf{Wn}^{E^*}\langle\Gamma\rangle = \top$. □

## 4 Soundness of CL2

**Lemma 4.1** *If* $\mathbf{CL2} \vdash F$, *then $F$ is valid (any formula $F$).*

*Moreover, there is an effective procedure that takes a $\mathbf{CL2}$-proof of an arbitrary formula $F$ and returns an HPM $\mathcal{H}$ such that, for every interpretation $^*$, $\mathcal{H} \models F^*$.*

**Proof.** Proposition 3.9 of [3] implies that, for any formula $F$ and interpretation $^*$, the game $F^*$ is static. Therefore, in view of Proposition 6.1 of [3], it would be sufficient to prove the above lemma — in particular, the 'Moreover' clause of it — with "fair EPM $\mathcal{E}$" instead of "HPM $\mathcal{H}$".

Furthermore, it would be sufficient to restrict interpretations to perfect ones. Indeed, suppose a machine $\mathcal{M}$ (whether it be an EPM or an HPM) wins $F^\dagger$ for every perfect interpretation $^\dagger$, and let $^*$ be a not-necessarily-perfect interpretation. We want to see that the same machine $\mathcal{M}$ also wins $F^*$. Suppose this is not the case, i.e. $\mathcal{M}$ loses $F^*$ on some input $e$. This means that, where $\Gamma$ is the run spelled by some $e$-computation branch of $\mathcal{M}$, we have $\mathbf{Wn}_e^{F^*}\langle\Gamma\rangle = \bot$. This means nothing but that $\mathbf{Wn}^{e[F^*]}\langle\Gamma\rangle = \bot$. Now, let $^\dagger$ be the perfect interpretation induced by $(^*, e)$. According to Lemma 3.2, $e[F^*] = F^\dagger$. Hence $\mathbf{Wn}^{F^\dagger}\langle\Gamma\rangle = \bot$, so that $\mathcal{M}$ does not win $F^\dagger$, which is a contradiction.

Finally, Lemma 3.1 allows us to safely replace "$\mathbf{CL2}$" by "$\mathbf{CL2}^\circ$" in our present lemma.

In view of the above observations, Lemma 4.1 is an immediate consequence of the following Lemma 4.2.
□

**Lemma 4.2** *There is an effective procedure that takes a $\mathbf{CL2}^\circ$-proof of an arbitrary formula $F$ and returns a fair EPM $\mathcal{E}$ such that, for every perfect interpretation $^*$, $\mathcal{E} \models F^*$.*

**Proof idea.** Every $\mathbf{CL2}^\circ$-proof, in fact, can be viewed as an input- and interpretation-independent winning strategy for $\top$, and the fair EPM $\mathcal{E}$ that we are going to design just follows such a strategy. As we probably remember from the soundness proof given in [3], the same was the case with $\mathbf{CL1}$, where each conclusion-to-premise transition of Rule **(a)** encoded a move by $\bot$ (with all premises accounting for all possible legal moves by $\bot$), and the conclusion-to-premise transition of Rule **(b)** encoded the "good" move that $\top$ should make in a given situation; in either case, after a move was made, $\top$'s strategy would "jump" to the corresponding premise $H$, recursively calling itself on $H$. In our present case this intuitive meaning of Rules **(a)** and **(b)** is retained. In addition, Rule **(c$^\circ$)** signals $\top$ that from now on it should try — using copy-cat methods — to keep identical[2] the subplays/subruns in the two occurrences of the hybrid atom introduced (in the bottom-up view) by that rule.

The overall situation with $\mathbf{CL1}$, however, was much simpler than it is with $\mathbf{CL2}^\circ$. In $\mathbf{CL1}$-proof-derived strategies, as we just noted, to every legal move in a play corresponded a transition from a given formula to one of its premises $H$ in the proof. This is no longer the case with $\mathbf{CL2}^\circ$. Specifically, there is nothing in $\mathbf{CL2}^\circ$-proofs corresponding to moves made in hybrid or general atoms. So, by the time when the strategy jumps to $H$, the game to which the original game will have been "brought down" may be not $H^*$ but rather $\langle\Omega\rangle H^*$, where $\Omega$ is the sequence of the moves made by the two players in the hybrid and general atoms of $H$. Thus, the strategy has to be successful for such $\langle\Omega\rangle H^*$ rather than (as this was the case with $\mathbf{CL1}$-proof-derived strategies) just for $H^*$. Fortunately, it turns out that success in this more complicated situation is still possible as long as $\Omega$ is $H$-manageable; and ensuring that $\Omega$ is indeed always manageable also turns out to be a "manageable" task for $\top$. This is where all of our manageability-related lemmas from the previous section come to help.

---

[2]This is generally impossible in the literal sense, but what *is* possible is to ensure that one play is a $\top$-delay of the other which, taking into account that we are talking about static games, is just as good as if the two plays were fully identical.

Now a little more detailed — yet informal — description of how our strategy/machine $\mathcal{E}$ for a $\mathbf{CL2}^\circ$-provable formula $F$ works. As noted, it is a recursive strategy, at every step dealing with $\langle\Omega\rangle E^*$ (* being irrelevant), where $E$ is a $\mathbf{CL2}^\circ$-provable hyperformula and $\Omega$ is an $E$-manageable (legal) position of $E^*$. Initially $E = F$ and $\Omega = \langle\rangle$. How $\mathcal{E}$ acts on $\langle\Omega\rangle E^*$ depends on by which of the three rules $E$ is derived in $\mathbf{CL2}^\circ$.

If $E$ is derived by Rule **(b)** from $H$, the machine — exactly as in [3] — makes the move $\alpha$ "prescribed" by that application of the rule. Say, if $E = (G_1 \sqcup G_2) \wedge (G_3 \sqcap G_4)$ and $H = G_2 \wedge (G_3 \sqcap G_4)$, then '1.2' is such a move. Lemma 3.11 tells us that $\Omega$ remains $H$-manageable and that $\alpha$ brings $\langle\Omega\rangle E^*$ down to $\langle\Omega\rangle H^*$. So, after making move $\alpha$, the machine switches to its winning strategy for $\langle\Omega\rangle H^*$. This, by the induction hypothesis, guarantees success.

If $E$ is derived by Rule **(c$^\circ$)** from $H$ through replacing the two occurrences of a hybrid atom $P_q$ in $H$ by $P$, then the machine finds within $\Omega$ and copies, in the positive occurrence of $P_q$, all of the moves made so far by the environment in the negative occurrence of $P_q$ (or rather in the corresponding occurrence of $P$), and vice versa. This series of moves brings the game down to $\langle\Omega'\rangle E^* = \langle\Omega'\rangle H^*$, where $\Omega'$ is result of adding those moves to $\Omega$. Lemma 3.13 guarantees that $\Omega'$ is $H$-manageable. So, now the machine switches to its successful strategy for $\langle\Omega'\rangle H^*$ and eventually wins.

Finally, suppose $E$ is derived by Rule **(a)**. Our machine keeps granting permission. Now and then the environment may be making moves in general atoms of $E$, to which $\mathcal{E}$ does not react. However, every time $\bot$ makes a move in one of the hybrid atoms, $\mathcal{E}$ copies that move in the other occurrence of the same hybrid atom. Clauses 1 and 2 of Lemma 3.14 guarantee that, while this is going on, (the continuously updated) $\Omega$ remains $E$-manageable. So, if nothing else happens, in view of Lemma 3.10, $\Omega$ — even if its grows infinite — remains $E$-manageable, and then Lemma 3.21 guarantees that the game will be won by $\top$ because, as a conclusion of Rule **(a)**, $E$ is stable. However, what will typically happen during this stage (except one — the last — case) is that sooner or later $\bot$ makes a legal move *not* in a hybrid or general atom, but rather a move signifying a choice associated with a $\sqcap$- or $\sqcup$-subformula of $E$. E.g., if $E = (P_q \vee \neg P_q) \wedge (G_3 \sqcap G_4)$, then '2.1' can be such a move. Now the situation is very similar to the case with Rule **(b)**: the machine simply switches to its winning strategy for $\langle\Omega\rangle H^*$, where $H$ is the corresponding premise of $E$ ($H = (P_q \vee \neg P_q) \wedge G_3$ in our example). Clause 3(b) of Lemma 3.14 guarantees that $\langle\Omega\rangle H^*$ is indeed the game to which $\langle\Omega\rangle E^*$ has evolved; and, according to clause 3(a) of the same lemma, $\Omega$ is $H$-manageable, so that, by the induction hypothesis, $\mathcal{E}$ knows how to win $\langle\Omega\rangle H^*$.

**Proof.** Fix a formula $F$ together with a $\mathbf{CL2}^\circ$-proof for it. In the present context we view such a proof as a sequence (rather than tree) of hyperformulas. We will be referring to this sequence as "the proof", and referring to the hyperformulas occurring in the proof as "proof hyperformulas". We assume that there are no repetitions or other redundancies in the proof (otherwise eliminate them), and that each proof hyperformula comes with a fixed *justification* — an indication of by which rule and from what premises the hyperformula was derived.

We construct the EPM $\mathcal{E}$ whose work can be described as follows. At the beginning, this machine creates two records: $E$ to hold a hyperformula, and $\Omega$ to hold a position. It initializes $E$ to $F$ and $\Omega$ to $\langle\rangle$. After this, $\mathcal{E}$ follows the following interactive algorithm MAIN LOOP. The description of this algorithm assumes that, at the beginning of each iteration of MAIN LOOP or INNER LOOP, the following condition is satisfied:

(*The value of*) $E$ *is a proof hyperformula.*

That this condition is always satisfied can be immediately seen from the description of the algorithm.

**Procedure** MAIN LOOP: Act depending on which of the three rules was used (last) to derive $E$ in the proof:

**Case of Rule (b):** Let $H$ be the premise of $E$ in the proof. $H$ is the result of substituting, in $E$, a certain negative (resp. positive) surface occurrence of a subformula $G_1 \sqcap \ldots \sqcap G_n$ (resp. $G_1 \sqcup \ldots \sqcup G_n$) by $G_i$ for some $i \in \{1, \ldots, n\}$. Let $\gamma$ be the $E$-specification of that occurrence. Then make the move $\gamma i$; update $E$ to $H$; repeat MAIN LOOP.

**Case of Rule (c°):** Let $H$ be the premise of $E$ in the proof. $H$ is the result of replacing in $E$ some positive surface occurrence $\pi$ and some negative surface occurrence $\nu$ of a general atom $P$ by a hybrid atom $P_q$. Let $\langle \bot\pi_1, \ldots, \bot\pi_n \rangle$ and $\langle \bot\nu_1, \ldots, \bot\nu_m \rangle$ be $\Omega^\pi$ and $\Omega^\nu$, respectively. Then: make the $m+n$ moves $\pi\nu_1, \ldots, \pi\nu_m, \nu\pi_1, \ldots, \nu\pi_n$ (in this very order); update $\Omega$ to $\langle \Omega, \top\pi\nu_1, \ldots, \top\pi\nu_m, \top\nu\pi_1, \ldots, \top\nu\pi_n \rangle$; update $E$ to $H$; repeat MAIN LOOP.

**Case of Rule (a):** Follow the procedure INNER LOOP described below.

INNER LOOP: Keep granting permission until the adversary makes a move $\alpha$, then act depending on which of the following four subcases holds:

**Subcase (i):** $\alpha = \gamma\beta$, where $\gamma$ $E$-specifies a surface occurrence of a general atom. Then update $\Omega$ to $\langle \Omega, \bot\gamma\beta \rangle$, and repeat INNER LOOP.

**Subcase (ii):** $\alpha = \gamma\beta$, where $\gamma$ $E$-specifies a surface occurrence of a hybrid atom. Let $\sigma$ be the $E$-specification of the other occurrence of the same hybrid atom. Then make the move $\sigma\beta$, update $\Omega$ to $\langle \Omega, \bot\gamma\beta, \top\sigma\beta \rangle$, and repeat INNER LOOP.

**Subcase (iii):** $\alpha = \gamma i$, where $\gamma$ $E$-specifies a positive (resp. negative) surface occurrence of a subformula $G_1 \sqcap \ldots \sqcap G_n$ (resp. $G_1 \sqcup \ldots \sqcup G_n$) and $i \in \{1, \ldots, n\}$. Let $H$ be the result of substituting in $E$ the above occurrence by $G_i$. Then update $E$ to $H$, and repeat MAIN LOOP.

**Subcase (iv):** $\alpha$ does not satisfy the conditions of any of the above Subcases (i),(ii),(iii). Then go to an infinite loop in a permission state.

It is obvious that (the description of) $\mathcal{E}$ can be constructed effectively from the **CL2°**-proof of $F$. What we need to do now is to show that $\mathcal{E}$ is fair and that it wins $F^*$ for every perfect interpretation $^*$.

Pick an arbitrary perfect interpretation $^*$, an arbitrary input $e$ (which is ignored by $\mathcal{E}$ anyway) and an arbitrary $e$-computation branch $B$ of $\mathcal{E}$. Fix $\Gamma$ as the run spelled by $B$. Consider the work of $\mathcal{E}$ in $B$. For each $k \geq 1$ such that MAIN LOOP makes at least $k$ iterations in $B$, let $E_k$ denote the value of the record $E$ at the beginning of the $k$th iteration of MAIN LOOP. Thus, $E_1 = F$. Since $^*$ is a perfect interpretation, for any hyperformula $H$, $H^*$ and $e[H^*]$ are the same. In particular, $e[F^*] = F^* = E_1^*$. Our goal is to show that $B$ is fair and $\mathbf{Wn}^{F^*}\langle\Gamma\rangle = \top$, i.e. $\mathbf{Wn}^{E_1^*}\langle\Gamma\rangle = \top$.

Evidently $E_{k+1}$ (as long as the $(k+1)$th iteration of MAIN LOOP exists) is always one of the premises of $E_k$ in the proof, so that MAIN LOOP is iterated only a finite number of times. Fix $l$ as the number of iterations of MAIN LOOP. The $l$th iteration deals with the case of Rule **(a)** — and, besides, never with Subcase (iii) within it — for otherwise there would be a next iteration. This guarantees that $\mathcal{E}$ will grant permission infinitely many times during the $l$th iteration, so that branch $B$ is indeed fair.

Thus, our remaining duty now is to show that $\mathbf{Wn}^{E_1^*}\langle\Gamma\rangle = \top$. By condition (c) of Definition 3.1 of [3], $\mathbf{Wn}^{E_1^*}\langle\Gamma\rangle = \top$ is immediate when $\Gamma$ is a $\bot$-illegal run of $E_1^*$. Hence we exclude this trivial case and, for the rest of this proof, assume that $\Gamma$ is not a $\bot$-illegal run of $E_1^*$. Speaking less formally, we assume that $\bot$ never makes illegal moves.

The fact that $E_l$ is derived by Rule **(a)** implies that

$$E_l \text{ is stable.} \tag{23}$$

For each $k$ with $1 \leq k \leq l$, let $\Theta_k$ be the sequence of the moves made by the players by the beginning of the $k$th iteration of MAIN LOOP, where the moves made by $\mathcal{E}$ are $\top$-labeled and the moves made by its adversary are $\bot$-labeled. Also, for each such $k$, let $\Omega_k$ be the value of record $\Omega$ at the beginning of the $k$th iteration of MAIN LOOP.

**Claim 1.** *For any $k$ with $1 \leq k \leq l$,*

$$\Omega_k \text{ is } E_k\text{-manageable;} \tag{24}$$
$$\langle\Theta_k\rangle E_1^* = \langle\Omega_k\rangle E_k^*. \tag{25}$$

This claim can be proven by induction on $k$. The basis case with $k=1$ is trivial as $\Theta_1 = \Omega_1 = \langle\rangle$.

Now consider an arbitrary $k$ with $1 \leq k < l$ and assume (induction hypothesis) that conditions (24)-(25) are satisfied. We separately consider the following three cases, depending on with which case the $k$th iteration of MAIN LOOP deals. In each case we want to show that the above two conditions continue to be satisfied for $k+1$, i.e. that the following statements are true:

$$\Omega_{k+1} \text{ is } E_{k+1}\text{-manageable;} \tag{26}$$

$$\langle \Theta_{k+1} \rangle E_1^* = \langle \Omega_{k+1} \rangle E_{k+1}^*. \tag{27}$$

**Case of Rule (b).** Record $\Omega$ is not updated in this case, so $\Omega_{k+1} = \Omega_k$. Exactly one ($\top$-labeled) move $\gamma i$ is made during the $k$th iteration of MAIN LOOP, where $\gamma$ is the $E_k$-specification of a negative (resp. positive) occurrence of a subformula $G_1 \sqcap \ldots \sqcap G_n$ (resp. $G_1 \sqcup \ldots \sqcup G_n$) of $E_k$ and $i \in \{1, \ldots, n\}$. Thus, $\Theta_{k+1} = \langle \Theta_k, \top\gamma i\rangle$. Also, $E_{k+1}$ relates to $E_k$ as $H$ does to $E$ in the description of the "Case of Rule **(b)**" part of MAIN LOOP and hence as in Lemma 3.11. Keeping this in mind, with $\Omega_k = \Omega_{k+1}$ in the role of $\Omega$, (26) follows from clause 1 of Lemma 3.11. According to clause 2 of the same lemma, $\langle \Omega_k, \top\gamma i\rangle E_k^* = \langle \Omega_k \rangle E_{k+1}^* = \langle \Omega_{k+1} \rangle E_{k+1}^*$, i.e., by Lemma 3.4,[3] $\langle \top\gamma i\rangle \langle \Omega_k \rangle E_k^* = \langle \Omega_{k+1} \rangle E_{k+1}^*$. But, by (25), $\langle \Omega_k \rangle E_k^* = \langle \Theta_k \rangle E_1^*$. Hence, $\langle \top\gamma i\rangle \langle \Theta_k \rangle E_1^* = \langle \Omega_{k+1} \rangle E_{k+1}^*$, i.e. $\langle \Theta_k, \top\gamma i\rangle E_1^* = \langle \Omega_{k+1} \rangle E_{k+1}^*$, which, as $\langle \Theta_k, \top\gamma i\rangle = \Theta_{k+1}$, proves (27).

**Case of Rule (c°).** With $E_k$ in the role of $E$ and $\Omega_k$ in the role of $\Omega$, let $H$, $\pi$, $\nu$, $\pi_1, \ldots, \pi_n$, $\nu_1, \ldots, \nu_m$ be as in the description of the 'Case of Rule **(c°)**' step of MAIN LOOP. Note that $E_{k+1} = H$ and $\Omega_{k+1} = \langle \Omega_k, \top\pi\nu_1, \ldots, \top\pi\nu_m, \top\nu\pi_1, \ldots, \top\nu\pi_n\rangle$. Now, with $E = E_k$ and $\Omega = \Omega_k$, the conditions of Lemma 3.13 are satisfied (in view of Convention 3.3, the condition $\langle \Omega_k \rangle \in \mathbf{LR}^{E_k}$ is implicitly contained in (25)). Therefore, by that lemma, $\Omega_{k+1}$ is an $E_{k+1}$-manageable legal position of $E_{k+1}^*$, which proves (26). $\langle \Omega_{k+1} \rangle E_{k+1}^*$ can be rewritten as $\langle \top\pi\nu_1, \ldots, \top\pi\nu_m, \top\nu\pi_1, \ldots, \top\nu\pi_n\rangle \langle \Omega_k \rangle E_{k+1}^*$. By (25), $\langle \Omega_k \rangle E_k^* = \langle \Theta_k \rangle E_1^*$. Also notice that $E_{k+1}^* = E_k^*$, so $\langle \Omega_k \rangle E_{k+1}^* = \langle \Theta_k \rangle E_1^*$. Hence $\langle \Omega_{k+1} \rangle E_{k+1}^* = \langle \top\pi\nu_1, \ldots, \top\pi\nu_m, \top\nu\pi_1, \ldots, \top\nu\pi_n\rangle \langle \Theta_k \rangle E_1^*$. This means that $\langle \Omega_{k+1} \rangle E_{k+1}^* = \langle \Theta_k, \top\pi\nu_1, \ldots, \top\pi\nu_m, \top\nu\pi_1, \ldots, \top\nu\pi_n\rangle E_1^*$. But obviously $\Theta_{k+1} = \langle \Theta_k, \top\pi\nu_1, \ldots, \top\pi\nu_m, \top\nu\pi_1, \ldots, \top\nu\pi_n\rangle$. Hence $\langle \Omega_{k+1} \rangle E_{k+1}^* = \langle \Theta_{k+1} \rangle E_1^*$. This proves (27).

**Case of Rule (a).** The work of the machine during this ($k$th) iteration of MAIN LOOP consists of iterating INNER LOOP a finite number $l_k \geq 1$ of times (otherwise we would have $k = l$). Fix this $l_k$. For each $m$ with $1 \leq m \leq l_k$, let $\Theta_{k,m}$ be the sequence of the (correspondingly labeled) moves made by the players at the beginning of the $m$th iteration of INNER LOOP within the $k$th iteration of MAIN LOOP. Thus, $\Theta_{k,1} = \Theta_k$. Similarly, for each such $m$, let $\Omega_{k,m}$ be the value of record $\Omega$ at the beginning of the $m$th iteration of INNER LOOP within the $k$th iteration of MAIN LOOP.

**Subclaim 1.1.** *For any $m$ with $1 \leq m \leq l_k$,*

$$\Omega_{k,m} \text{ is } E_k\text{-manageable;} \tag{28}$$

$$\langle \Theta_{k,m} \rangle E_1^* = \langle \Omega_{k,m} \rangle E_k^*. \tag{29}$$

We prove the above by induction on $m$. The basis case with $m = 1$ is straightforward: we have $\Theta_{k,1} = \Theta_k$ and $\Omega_{k,1} = \Omega_k$. Hence the basis of our induction on $m$ is nothing but the induction hypothesis of the induction on $k$ in our proof of Claim 1. That is, it is nothing but statements (24)-(25).

Now consider any $m$ with $1 \leq m < l_k$. Assume statements (28)-(29) are true (induction hypothesis). We want to show that then the following conditions are also satisfied:

$$\Omega_{k,m+1} \text{ is } E_k\text{-manageable;} \tag{30}$$

$$\langle \Theta_{k,m+1} \rangle E_1^* = \langle \Omega_{k,m+1} \rangle E_k^*. \tag{31}$$

At the beginning of the $m$th iteration of INNER LOOP, the machine is waiting for the adversary to make a move $\alpha$. Such a move must be made because $k$ is not the last iteration of MAIN LOOP. By our assumption that $\bot$ never makes illegal moves (together the assumption that $\Theta_{k,m} \in \mathbf{LR}^{E_1^*}$ which is implied by (29)), $\langle \Theta_{k,m}, \bot\alpha \rangle$ must be a legal position of $E_1^*$, whence $\langle \bot\alpha \rangle$ is a legal position of $\langle \Theta_{k,m} \rangle E_1^*$, whence, by (29), $\langle \bot\alpha \rangle$ is a legal position of $\langle \Omega_{k,m} \rangle E_k^*$, i.e. $\langle \Omega_{k,m}, \bot\alpha \rangle$ is a legal position of $E_k^*$. This means that $\alpha$, $E_k$ (in the role of $E$) and $\Omega_{k,m}$ (in the role of $\Omega$) satisfy the conditions of Lemma 3.14. Then, in view of that lemma, it is obvious that $\alpha$ should satisfy (the conditions of) one of the Subcases **(i)**,**(ii)** or **(iii)** from the description of INNER LOOP. Subcase **(iii)** is impossible because then we would have $m = l_k$. Thus, either Subcase **(i)** or Subcase **(ii)** should be satisfied.

---

[3]Henceforth we will be using Lemma 3.4 without explicitly mentioning it.

Suppose Subcase **(i)** is satisfied, with $\alpha = \gamma\beta$ as described in that subcase. Then $\Omega_{k,m+1} = \langle \Omega_{k,m}, \bot\gamma\beta \rangle$, and (30) holds by clause 1 of Lemma 3.14, which also asserts that $\langle \Omega_{k,m}, \bot\gamma\beta \rangle$ is a legal position of $E_k^*$. According to (29), $\langle \Omega_{k,m} \rangle E_k^* = \langle \Theta_{k,m} \rangle E_1^*$, and hence $\langle \Omega_{k,m}, \bot\gamma\beta \rangle E_k^* = \langle \Theta_{k,m}, \bot\gamma\beta \rangle E_1^*$, i.e. $\langle \Omega_{k,m+1} \rangle E_k^* = \langle \Theta_{k,m}, \bot\gamma\beta \rangle E_1^*$. But in our case $\Theta_{k,m+1} = \langle \Theta_{k,m}, \bot\gamma\beta \rangle$. Therefore, $\langle \Theta_{k,m+1} \rangle E_1^* = \langle \Omega_{k,m+1} \rangle E_k^*$. This proves (31).

Suppose now Subcase **(ii)** is satisfied, with $\alpha = \gamma\beta$ and $\sigma$ as described in that subcase. Then $\Omega_{k,m+1} = \langle \Omega_{k,m}, \bot\gamma\beta, \top\sigma\beta \rangle$. Now, (30) follows from clause 2 of Lemma 3.14, which also asserts that $\langle \Omega_{k,m}, \bot\gamma\beta, \top\sigma\beta \rangle \in \mathbf{LR}^{E_k^*}$. And, taking into account that $\Theta_{k,m+1} = \langle \Theta_{k,m}, \bot\gamma\beta, \top\sigma\beta \rangle$, (31) follows from (29). Subclaim 1.1 is proven.

Back to the 'Case of Rule **(a)**' step of our proof of Claim 1. Consider the $l_k$th (last) iteration of INNER LOOP within the $k$th iteration of MAIN LOOP. Since $k < l$, obviously this iteration deals with Subcase **(iii)**. Let $\alpha$ and $H$ be as in the description of that subcase. Note that $E_{k+1} = H$ and $\Omega_{k+1} = \Omega_{k,l_k}$. For the same reasons as in the proof of Subclaim 1.1, $\langle \Omega_{k,l_k}, \bot\alpha \rangle$ is a legal position of $E_k^*$. According to Subclaim 1.1, we have:

$$\Omega_{k,l_k} \text{ is } E_k\text{-manageable;} \tag{32}$$

$$\langle \Theta_{k,l_k} \rangle E_1^* = \langle \Omega_{k,l_k} \rangle E_k^*. \tag{33}$$

Statement (26) follows from (32) by clause 3(a) of Lemma 3.14. According to clause 3(b) of the same lemma, $\langle \Omega_{k,l_k}, \bot\alpha \rangle E_k^* = \langle \Omega_{k,l_k} \rangle E_{k+1}^*$ and hence $\langle \Omega_{k,l_k}, \bot\alpha \rangle E_k^* = \langle \Omega_{k+1} \rangle E_{k+1}^*$. This, in turn, implies $\langle \bot\alpha \rangle \langle \Omega_{k,l_k} \rangle E_k^* = \langle \Omega_{k+1} \rangle E_{k+1}^*$. By (33), $\langle \Omega_{k,l_k} \rangle E_k^* = \langle \Theta_{k,l_k} \rangle E_1^*$. Hence $\langle \bot\alpha \rangle \langle \Theta_{k,l_k} \rangle E_1^* = \langle \Omega_{k+1} \rangle E_{k+1}^*$ and thus $\langle \Theta_{k,l_k}, \bot\alpha \rangle E_1^* = \langle \Omega_{k+1} \rangle E_{k+1}^*$. But observe that $\langle \Theta_{k,l_k}, \bot\alpha \rangle = \Theta_{k+1}$. Therefore (27) holds.

Claim 1 is proven.

We continue our proof of Lemma 4.2. Consider the last ($l$th) iteration of MAIN LOOP. As we noted earlier when deriving (23), this iteration deals with the case of Rule **(a)**. Let $\mathcal{N}$ be $\{1, \ldots, k\}$ if $k$ is the number of iterations of INNER LOOP within the $l$th iteration of MAIN LOOP, and be $\{1, 2, 3, \ldots\}$ if there are infinitely many such iterations. For each $m \in \mathcal{N}$, as before, let $\Theta_{l,m}$ be the sequence of the correspondingly labeled moves made in the overall run by the beginning of the $m$th iteration of INNER LOOP within the $l$th iteration of MAIN LOOP, and let $\Omega_{l,m}$ be the value of record $\Omega$ at the beginning of the $m$th iteration of INNER LOOP within the $l$th iteration of MAIN LOOP. For the same reasons[4] as in the proof of (28) and (29) (where from (29) we only need its implicit statement that $\Omega_{k,m} \in \mathbf{LR}^{E_k^*}$), we have:

$$\text{For any } m \in \mathcal{N}, \ \Omega_{l,m} \text{ is an } E_l\text{-manageable legal position of } E_l^*. \tag{34}$$

Note that, during the work of $\mathcal{E}$, every update of record $\Omega$ extends its previous value by adding new labeled moves to it, without ever deleting old labeled moves. So, let $\Omega_\infty$ be the "ultimate" value of $\Omega$, precisely meaning the shortest run such that, for every $m \in \mathcal{N}$, $\Omega_{l,m}$ is an initial segment of $\Omega_\infty$. Of course, if $\mathcal{N} = \{1, \ldots, k\}$, then $\Omega_\infty$ is simply $\Omega_{l,k}$. Statement (34) — together with Lemma 3.10 when $\Omega_\infty$ is infinite — implies that $\Omega_\infty$ is an $E_l$-manageable legal run of $E_l^*$. Therefore, by (23) and Lemma 3.21, we have

$$\mathbf{Wn}^{E_l^*}\langle \Omega_\infty \rangle = \top. \tag{35}$$

$\Omega_\infty$ is an extension of $\Omega_l$, so that $\Omega_\infty = \langle \Omega_l, \Delta \rangle$ for some run $\Delta$. Let us fix this $\Delta$. Now (35) can be rewritten as $\mathbf{Wn}^{E_l^*}\langle \Omega_l, \Delta \rangle = \top$. In turn, the latter — remembering the definition of the operation of prefixation and taking into account that, by (25), $\Omega_l \in \mathbf{LR}^{E_l^*}$ — can be rewritten as $\mathbf{Wn}^{\langle \Omega_l \rangle E_l^*}\langle \Delta \rangle = \top$. Now, according to Claim 1, $\langle \Omega_l \rangle E_l^* = \langle \Theta_l \rangle E_1^*$. Thus, $\mathbf{Wn}^{\langle \Theta_l \rangle E_1^*}\langle \Delta \rangle = \top$, which can be rewritten back as

$$\mathbf{Wn}^{E_1^*}\langle \Theta_l, \Delta \rangle = \top. \tag{36}$$

---

[4]With the minor difference that now the reason why Subcase (iii) is impossible is that otherwise $l$ would not be the last iteration of MAIN LOOP.

For the same reasons[5] as in our proof of Subclaim 1, the $l$th iteration of MAIN LOOP never deals with Subcases (iii) or (iv) of 'Case of Rule **(a)**'. The remaining Subcases (i) and (ii) add to record $\Omega$ all of the moves made by the players (and no other moves, of course). Therefore, taking into account that the value of that record is $\Omega_l$ when the $l$th iteration of INNER LOOP starts, we can see that $\Delta$ is nothing but exactly the sequence of all moves made during the $l$th iteration of INNER LOOP. Hence $\Gamma = \langle \Theta_l, \Delta \rangle$ where, as we remember, $\Gamma$ is the run spelled by the computation branch $B$ of $\mathcal{E}$ that we are considering. Thus, by (36), $\mathbf{Wn}^{E_1^*}\langle \Gamma \rangle = \top$, and our proof of Lemma 4.2 is complete. □

## 5 Completeness of CL2

**Lemma 5.1** *If* $\mathbf{CL2} \not\vdash F$, *then $F$ is not valid (any formula $F$).*

*Moreover, if* $\mathbf{CL2} \not\vdash F$, *then $F^*$ is not computable for some interpretation $*$ that interprets all elementary atoms of $F$ as finitary predicates of arithmetical complexity $\Delta_2$, and interprets all general atoms of $F$ as problems of the form $(A_1^1 \sqcup \ldots \sqcup A_m^1) \sqcap \ldots \sqcap (A_1^m \sqcup \ldots \sqcup A_m^m)$, where each $A_i^j$ is a finitary predicate of arithmetical complexity $\Delta_2$.*

**Proof idea**. We are going to show that if $\mathbf{CL2} \not\vdash F$, then there is an elementary-base formula $\lceil F \rceil$ of the same form as $F$ that is not provable in $\mathbf{CL1}$. Precisely, "the same form as $F$" here means that $\lceil F \rceil$ is the result of rewriting/expanding in $F$ every general atom $P$ as a certain elementary-base formula $\check{P}_\sqcup^\sqcap$. This, in view of the already known completeness of $\mathbf{CL1}$, immediately yields non-validity for $F$. As it turns out, the above formulas $\check{P}_\sqcup^\sqcap$ can be chosen to be as simple as sufficiently long $\sqcap$-conjunctions of sufficiently long $\sqcup$-disjunctions of arbitrary "neutral" (not occurring in $F$ and pairwise distinct) elementary atoms, with the "sufficient length" of those conjuncts/disjuncts being bounded by the number of occurrences of general atoms in $F$.

Intuitively, the reason why $\mathbf{CL1} \not\vdash \lceil F \rceil$, i.e. why $\top$ cannot win (the game represented by) $\lceil F \rceil$, is that a smart environment may start choosing different conjuncts/disjuncts in different occurrences of $\check{P}_\sqcup^\sqcap$. The best that $\top$ can do in such a play is to match any given positive or negative occurrence of $\check{P}_\sqcup^\sqcap$ with one (but not more!) negative or positive occurrence of the same subgame — match in the sense that mimic environment's moves in order to keep the subgames/subformulas at the two occurrences identical. Yet, this is insufficient for $\top$ to achieve a guaranteed success. For, if it was sufficient, then every decision about what to match with what in $\lceil F \rceil$ could be modeled, in an attempted $\mathbf{CL2}$-proof for $F$, by an appropriate application of Rule **(c)**; this, together with the possibility to model — through Rules **(a)** and **(b)** — $\bot$'s and $\top$'s decisions required by the choice connectives in the "ordinary" (non-$\check{P}_\sqcup^\sqcap$) parts of $\lceil F \rceil$, would eventually make $F$ $\mathbf{CL2}$-provable, which however it is not.

**Proof.** Fix a formula $F$. Let $\mathcal{P}$ be the set of all general atoms occurring in $F$. Let us fix $m$ as the total number of occurrences of such atoms in $F$;[6] if there are fewer than 2 of such occurrences, then we take $m = 2$.

For the rest of this section, let us agree that

$$a, b \text{ always range over } \{1, \ldots, m\}.$$

For each $P \in \mathcal{P}$ and each $a, b$, let us fix an elementary atom

- $\check{P}_b^a$

---

[5] Again, with the minor difference pointed out in the previous footnote.

[6] In fact, a much smaller $m$ would be sufficient for our purposes. E.g., $m$ can be chosen to be such that no given general atom has more than $m$ occurrences in $F$. But why try to economize?

not occurring in $F$. We assume that $\check{P}^a_b \neq \check{Q}^c_d$ as long as either $P \neq Q$ or $a \neq c$ or $b \neq d$. Note that the $\check{P}^a_b$ are elementary atoms despite our "tradition" according to which the capital letters $P, Q, \ldots$ stand for general atoms.

Next, for each $P \in \mathcal{P}$ and each $a$, we define

- $\check{P}^a_\sqcup = \check{P}^a_1 \sqcup \ldots \sqcup \check{P}^a_m$.

Finally, for each $P \in \mathcal{P}$, we define

- $\check{P}^\sqcap_\sqcup = \check{P}^1_\sqcup \sqcap \ldots \sqcap \check{P}^m_\sqcup$, i.e. $\check{P}^\sqcap_\sqcup = (\check{P}^1_1 \sqcup \ldots \sqcup \check{P}^1_m) \sqcap \ldots \sqcap (\check{P}^m_1 \sqcup \ldots \sqcup \check{P}^m_m)$.

We refer to the above formulas $\check{P}^a_b$, $\check{P}^a_\sqcup$ and $\check{P}^\sqcap_\sqcup$ as **molecules**, in particular, **$P$-based molecules**. To differentiate between the three sorts of molecules, we call the molecules of the type $\check{P}^a_b$ **small**, call the molecules of the type $\check{P}^a_\sqcup$ **medium**, and call the molecules of the type $\check{P}^\sqcap_\sqcup$ **large**. Thus, where $k$ is the cardinality of $\mathcal{P}$, altogether there are $k$ large molecules, $k \times m$ medium molecules and $k \times m \times m$ small molecules.

For simplicity, for the rest of this section we assume/pretend that the languages of **CL1** and **CL2** have no nonlogical atoms other than those occurring in $F$ plus the atoms $\check{P}^b_a$ ($P \in \mathcal{P}$, $a, b \in \{1, \ldots, m\}$). This way the scopes of the terms "formula" (meaning formula of the language **CL2**) and "elementary-base formula" (meaning formula of **CL1**) are correspondingly redefined.

Let us say that an occurrence of a molecule in a given elementary-base formula is **independent** iff it is not a part of another ("larger") molecule. E.g., the occurrence of $\check{P}^a_b$ in $\check{P}^a_b \to \bot$ is independent, while in $\check{P}^a_\sqcup \to \bot$, i.e. in $\check{P}^a_1 \sqcup \ldots \sqcup \check{P}^a_b \sqcup \ldots \sqcup \check{P}^a_m \to \bot$, it is not. Of course, surface occurrences of molecules are always independent, and so are any — surface or non-surface — occurrences of large molecules.

We say that an elementary-base formula $E$ is **good** iff the following conditions are satisfied:

**Cond1:** $E$ contains at most $m$ independent occurrences of molecules.

**Cond2:** Only large molecules (may) have independent non-surface occurrences in $E$.

**Cond3:** Each small molecule has at most one positive and at most one negative independent occurrence in $E$.

**Cond4:** For each medium molecule $\check{P}^a_\sqcup$, $E$ has at most one positive independent occurrence of $\check{P}^a_\sqcup$, and when $E$ has such an occurrence, then for no $b$ does $E$ have a positive independent occurrence of the small molecule $\check{P}^a_b$.

Let $E$ be an elementary-base formula. By an **isolated** small molecule of $E$ (or $E$-**isolated** small molecule, or a small molecule **isolated in** $E$) we will mean a small molecule that has exactly one independent occurrence in $E$; we will say that such a molecule is **positive** or **negative** depending on whether its independent occurrence in $E$ is positive or negative. Next, the **floorification** of $E$, denoted

$$\lfloor E \rfloor,$$

is the result of replacing in $E$ every independent occurrence of every $P$-based (each $P \in \mathcal{P}$) large, medium and $E$-isolated small molecule by the general atom $P$.

**Claim 1.** *For any good elementary-base formula $E$, if $\mathbf{CL1} \vdash E$, then $\mathbf{CL2} \vdash \lfloor E \rfloor$.*

To prove this claim, assume $E$ is a good elementary-base formula, and $\mathbf{CL1} \vdash E$. By induction on the length of the **CL1**-proof of $E$, we want to show that $\mathbf{CL2} \vdash \lfloor E \rfloor$. We need to consider the following two cases, depending on which of the two rules of **CL1** was used (last) to derive $E$ in **CL1**.

*Case 1:* $E$ is derived by Rule **(a)**. Let us fix the set $\vec{H}$ of premises of $E$. Each formula $H \in \vec{H}$ is provable in **CL1**. Hence, by the induction hypothesis, we have:

$$\text{For any } H \in \vec{H}, \text{ if } H \text{ is good, then } \mathbf{CL2} \vdash \lfloor H \rfloor. \tag{37}$$

We consider the following 3 subcases. The first two subcases are not mutually exclusive, and either one can be chosen when both of them apply.

*Subcase 1.1:* $E$ has a positive surface occurrence of a large molecule $\check{P}_{\sqcup}^{\sqcap}$. Pick any $a$ such that neither the medium molecule $\check{P}_{\sqcup}^{a}$ nor any small molecule $\check{P}_{b}^{a}$ (whatever $b$) have independent occurrences in $E$. Such an $a$ exists, for otherwise we would have at least $m+1$ occurrences of molecules in $E$ (including the occurrence of $\check{P}_{\sqcup}^{\sqcap}$), which contradicts **Cond1**. Let $H$ be the result of replacing in $E$ the above occurrence of $\check{P}_{\sqcup}^{\sqcap}$ by $\check{P}_{\sqcup}^{a}$. Clearly $H \in \vec{H}$. Observe that when transferring from $E$ to $H$, we just "downsize" $\check{P}_{\sqcup}^{\sqcap}$ and otherwise do not create any additional independent occurrences of molecules, so **Cond1** continues to be satisfied for $H$. Neither do we introduce any new non-surface occurrences of molecules or any new independent occurrences of small molecules, so **Cond2** and **Cond3** also continue to hold for $H$. And our choice of $a$ obviously guarantees that so does **Cond4**. To summarize, $H$ is good. Therefore, by (37), $\mathbf{CL2} \vdash \lfloor H \rfloor$. Finally, note that, when floorifying a given formula, both $\check{P}_{\sqcup}^{\sqcap}$ and $\check{P}_{\sqcup}^{a}$ get replaced by the same atom $P$; and, as the only difference between $E$ and $H$ is that $H$ has $\check{P}_{\sqcup}^{a}$ where $E$ has $\check{P}_{\sqcup}^{\sqcap}$, obviously $\lfloor H \rfloor = \lfloor E \rfloor$. Thus, $\mathbf{CL2} \vdash \lfloor E \rfloor$.

*Subcase 1.2:* $E$ has a negative surface occurrence of a medium molecule $\check{P}_{\sqcup}^{a}$. Pick any $b$ such that $E$ does not have an independent occurrence of $\check{P}_{b}^{a}$. Again, in view of **Cond1**, such a $b$ exists. Let $H$ be the result of replacing in $E$ the above occurrence of $\check{P}_{\sqcup}^{a}$ by $\check{P}_{b}^{a}$. Certainly $H \in \vec{H}$. Conditions **Cond1** and **Cond2** continue to hold for $H$ for the same reasons as in Subcase 1.1. In view of our choice of $b$, **Cond3** is also inherited by $H$ from $E$. And so is **Cond4**, because $H$ has the same positive occurrences of (the same) molecules as $E$ does. Thus, $H$ is good. Therefore, by (37), $\mathbf{CL2} \vdash \lfloor H \rfloor$. It remains to show that $\lfloor H \rfloor = \lfloor E \rfloor$. Note that when floorifying $E$, $\check{P}_{\sqcup}^{a}$ gets replaced by $P$. But so does $\check{P}_{b}^{a}$ when floorifying $H$ because, by our choice of $b$, $\check{P}_{b}^{a}$ is an isolated small molecule of $H$. Since the only difference between $H$ and $E$ is that $H$ has $\check{P}_{b}^{a}$ where $E$ has $\check{P}_{\sqcup}^{a}$, it is then obvious that indeed $\lfloor H \rfloor = \lfloor E \rfloor$.

*Subcase 1.3:* None of the above two conditions is satisfied. This means that in $E$ all surface occurrences of large molecules are negative, and all surface occurrences of medium molecules are positive. Every large molecule $\check{P}_{\sqcup}^{\sqcap}$ is a $\sqcap$-formula whose surface occurrences, as we remember, get replaced by $\top$ when transferring from $E$ to $\|E\|$; but the same happens to the corresponding occurrences of $P$ in $\lfloor E \rfloor$ when transferring from $\lfloor E \rfloor$ to $\|\lfloor E \rfloor\|$ because, as we have just noted, such occurrences are negative, and negative surface occurences of general atoms get replaced by $\top$ when elementarizing formulas. Similarly, every medium molecule $\check{P}_{\sqcup}^{a}$ is a $\sqcup$-formula so that its surface occurrences get replaced by $\bot$ when transferring from $E$ to $\|E\|$; but the same happens to the corresponding occurrences of $P$ in $\lfloor E \rfloor$ when transferring from $\lfloor E \rfloor$ to $\|\lfloor E \rfloor\|$ because they are positive, and positive surface occurences of general atoms get replaced by $\bot$ when elementarizing formulas. Based on these observations, with a little thought we can see that $\|\lfloor E \rfloor\|$ is "almost the same" as $\|E\|$; specifically, the only difference between these two formulas is that $\|\lfloor E \rfloor\|$ has $\bot$ where $\|E\|$ has positive isolated small molecules, and $\|\lfloor E \rfloor\|$ has $\top$ where $\|E\|$ has negative isolated small molecules. Obviously this means that $\|\lfloor E \rfloor\|$ is a substitutional instance of $\|E\|$ — the result of substituting, in the latter, every positive isolated small molecule by $\bot$ and every negative isolated small molecule by $\top$. As $E$ is derived by Rule **(a)**, $\|E\|$ is classically valid. Therefore $\|\lfloor E \rfloor\|$, as a substitutional instance of $\|E\|$, is also classically valid. So, we have:

$$\lfloor E \rfloor \text{ is stable.} \qquad (38)$$

Now consider an arbitrary formula $H'$ that is the result of replacing in $\lfloor E \rfloor$ a positive (resp. negative) surface occurrence $\gamma$ of a subformula $G_1' \sqcap \ldots \sqcap G_n'$ (resp. $G_1' \sqcup \ldots \sqcup G_n'$) by $G_i'$ for some $i \in \{1, \ldots, n\}$. Our goal is to show that $\mathbf{CL2} \vdash H'$. If we succeed, then, in view of (38), we can conclude that $\lfloor E \rfloor$ is derivable in $\mathbf{CL2}$ by Rule **(a)**. The logical structure of $E$ is the same as that of $\lfloor E \rfloor$, with the only difference that, wherever $\lfloor E \rfloor$ has general atoms, $E$ has molecules. Hence the same $\gamma$ also $E$-specifies a positive (resp. negative) occurrence of a subformula $G_1 \sqcap \ldots \sqcap G_n$ (resp. $G_1 \sqcup \ldots \sqcup G_n$) of $E$. Let then $H$ be the result of replacing in $E$ this subformula by $G_i$. Of course $H \in \vec{H}$. So, in view of (37), all what would suffice to show (in order to find $\mathbf{CL2} \vdash H'$) is that $H$ is good and $H' = \lfloor H \rfloor$. Let us first see that $H$ is good. When transferring from $E$ to $H$, **Cond1** is inherited by $H$ for the same or a similar reason as in all of the previous cases. So is **Cond2** because we are not creating any new non-surface occurrences. Furthermore, notice that $G_1 \sqcap \ldots \sqcap G_n$ (resp. $G_1 \sqcup \ldots \sqcup G_n$) is not a molecule, for otherwise in $\lfloor E \rfloor$ we would have a general atom at $\gamma$ rather than $G'_1 \sqcap \ldots \sqcap G'_n$ (resp. $G'_1 \sqcup \ldots \sqcup G'_n$). Hence, in view of **Cond2**, $G_i$ is not a small or medium molecule. This means that, when transferring from $E$ to $H$, we are not creating new (nor destroying old) independent/surface occurrences of any small or medium molecules, so that **Cond3** and **Cond4** are also inherited by $H$ from $E$. To summarize, $H$ is indeed good. Finally, it is also rather obvious that $H' = \lfloor H \rfloor$. The only case when we might have $H' \neq \lfloor H \rfloor$ would be if there was a small molecule $\check{P}^a_b$ isolated in $E$ but not in $H$, or vice versa (so that the independent occurrence of that molecule in $E$ would become $P$ in $\lfloor E \rfloor$ and hence in $H'$ but stay $\check{P}^a_b$ in $\lfloor H \rfloor$, or vice versa). But, as we observed just a while ago, $E$ and $H$ do not differ in what independent/surface occurrences of what small molecules they have.

*Case 2:* $E$ is derived by Rule **(b)**. That is, we have **CL1** $\vdash H$, where $H$ is the result of replacing in $E$ a negative (resp. positive) surface occurrence of a quasiatom $G$ of the form $G_1 \sqcap \ldots \sqcap G_n$ (resp. $G_1 \sqcup \ldots \sqcup G_n$) by $G_i$ for some $i \in \{1, \ldots, n\}$. Fix these formulas and this number $i$. Just as in Case 1 (statement (37)), based on the induction hypothesis, we find:

$$\text{If } H \text{ is good, then } \mathbf{CL2} \vdash \lfloor H \rfloor. \tag{39}$$

We need to consider the following three subcases that cover all possibilities:

*Subcase 2.1:* $G$ is not a molecule. Reasoning (almost) exactly as we did at the end of our discussion of Subcase 1.3, we find that $H$ is good. Therefore, by (39), **CL2** $\vdash \lfloor H \rfloor$. Now, a little thought can convince us that $\lfloor E \rfloor$ follows from $\lfloor H \rfloor$ by Rule **(b)**, so that **CL2** $\vdash \lfloor E \rfloor$.

*Subcase 2.2:* $G$ is a large molecule $\check{P}^\sqcap_\sqcup$. So, the occurrence of $G$ in $E$ is negative, and $G_i = \check{P}^i_\sqcup$. A (now already routine for us) examination of **Cond1**-**Cond4** reveals that each of these four conditions are inherited by $H$ from $E$, so that $H$ is good. Therefore, by (39), **CL2** $\vdash \lfloor H \rfloor$. Now, $\lfloor H \rfloor$ can be easily seen to be the same as $\lfloor E \rfloor$, and thus **CL2** $\vdash \lfloor E \rfloor$.

*Subcase 2.3:* $G$ is a medium molecule $\check{P}^a_\sqcup$. So, the occurrence of $G$ in $E$ is positive, and $G_i = \check{P}^a_i$. There are two subsubcases to consider:

*Subsubcase 2.3.1:* $E$ contains no independent occurrence of $\check{P}^a_i$. One can easily verify that $H$ is good and that $\lfloor H \rfloor = \lfloor E \rfloor$. By (39), we then get the desired **CL2** $\vdash \lfloor E \rfloor$.

*Subsubcase 2.3.2:* $E$ has an independent occurrence of $\check{P}^a_i$. Since $E$ also has a positive independent occurrence of $\check{P}^a_\sqcup$, **Cond4** implies that the above occurrence of $\check{P}^a_i$ in $E$ is negative. This, in conjunction with **Cond3**, means that $E$ does not have any other independent occurrences of $\check{P}^a_i$, and thus $H$ has exactly two — one negative and one positive — independent occurrences of $\check{P}^a_i$. This guarantees that **Cond3** is satisfied for $H$, because $H$ and $E$ only differ in that $H$ has $\check{P}^a_i$ where $E$ has $\check{P}^a_\sqcup$. The conditions **Cond1** and **Cond2** are straightforwardly inherited by $H$ from $E$. Finally, **Cond4** also transfers from $E$ to $H$ because, even though $H$ — unlike $E$ — has a positive independent occurrence of $\check{P}^a_i$, it no longer has a positive independent occurrence of $\check{P}^a_\sqcup$ (which, by the same condition **Cond4** for $E$, was unique in $E$). Thus, $H$ is good and, by (39), **CL2** $\vdash \lfloor H \rfloor$. Note that since $H$ is good, by **Cond2**, both of the independent occurrences of $\check{P}^a_i$ in it are surface occurrences. The same, of course, is true for the corresponding occurrences of $\check{P}^a_i$ and $\check{P}^a_\sqcup$ in $E$. Let us now compare $\lfloor E \rfloor$ with $\lfloor H \rfloor$. According to our earlier observation, $\check{P}^a_j$ only has one independent occurrence in $E$, i.e. $\check{P}^a_i$ is $E$-isolated. Hence the independent occurrence of $\check{P}^a_i$, just as that of $\check{P}^a_\sqcup$, gets replaced by $P$ when floorifying $E$. On the other hand, $\check{P}^a_i$ is no longer isolated in $H$, so the two independent occurrences of it stay as they are when floorifying $H$. Based on this observation, we can easily see that the only difference between $\lfloor E \rfloor$ and $\lfloor H \rfloor$ is that $\lfloor E \rfloor$ has the general atom $P$ where $\lfloor H \rfloor$ has the (two occurrences of) elementary atom $\check{P}^a_i$. Since $\lfloor E \rfloor$ does not contain $\check{P}^a_i$ (because the only independent occurrence of it in $E$, as well as all large and medium $P$-based molecules, got replaced by $P$ when floorifying $E$), and since we are talking about two — one positive and one negative — surface occurrences of $P$ in $\lfloor E \rfloor$, we find that $\lfloor E \rfloor$ follows from $\lfloor H \rfloor$ by Rule **(c)**. We already know that **CL2** $\vdash \lfloor H \rfloor$. Hence **CL2** $\vdash \lfloor E \rfloor$.

Claim 1 is proven.

Now we are very close to finishing our proof of Lemma 5.1. Assume **CL2** $\nvdash F$. Let $\lceil F \rceil$ be the result of replacing in $F$ all occurrences of each general atom $P \in \mathcal{P}$ by $\check{P}^\sqcap_\sqcup$. Obviously $\lceil F \rceil$ is good. Clearly we also have $\lfloor \lceil F \rceil \rfloor = F$, so that **CL2** $\nvdash \lfloor \lceil F \rceil \rfloor$. Therefore, by Claim 1, **CL1** $\nvdash \lceil F \rceil$. Hence, by Lemma 8.2 and Remark 8.3 of [3], there is an interpretation $\dagger$ that interprets every elementary atom as a finitary predicate of arithmetical complexity $\Delta_2$, such that

$$\not\models \lceil F \rceil^{\dagger}. \tag{40}$$

Let $^*$ be an interpretation such that:

- $^*$ agrees with $^{\dagger}$ on all elementary atoms;

- $^*$ interprets each atom $P \in \mathcal{P}$ as $(\check{P}_{\sqcup}^{\sqcup})^{\dagger}$.

Clearly $^*$ interprets atoms as promised in our Lemma 5.1. It is also obvious that $F^* = \lceil F \rceil^{\dagger}$. Therefore, by (40), $\not\models F^*$, and the lemma is proven. $\square$

# References

[1] A. Blass, *A game semantics for linear logic.* **Annals of Pure and Applied Logic** 56 (1992), pp.183-220.

[2] G. Japaridze, *Introduction to computability logic.* **Annals of Pure and Applied Logic** 123 (2003), pp.1-99.

[3] G. Japaridze, *Propositional computability logic I.* **Transactions on Computational Logic** (to appear). A prepublication version is available at http://arxiv.org/abs/cs.LO/0404023.

[4] G. Japaridze, *Computability logic: a formal theory of interaction.* arXiv:cs.LO/0404024 (April 2004), 26 pages. URL: http://arxiv.org/abs/cs.LO/0404024.

# Index